\documentclass[aps,preprint]{revtex4-1}
\usepackage{amsmath}
\usepackage{amssymb}
\newtheorem{thm}{Theorem}[section]
\newtheorem{lem}[thm]{Lemma}
\newcommand{\R}{\mathbb{R}}

\begin{document}

\title[Vortex Equations Governing the FQHE]{Vortex Equations Governing the Fractional Quantum Hall Effect}

\author{Luciano Medina}
\email{lmedina@nyu.edu}
\affiliation{Department of Mathematics\\ 
Polytechnic School of Engineering\\ 
New York University\\
Brooklyn, NY 11201
USA}

%\date{\today}
%----------classification, keywords, date
%\subjclass{35J20, 35J47, 37K40, 81T13.}
\keywords{Chern-Simons theory, quantum Hall effect, vortices, Bogomol'nyi equations, calculus of variations, Sobolev Inequalities}

%%% ----------------------------------------------------------------------

\begin{abstract}
An existence theory is established for a coupled non-linear elliptic system, known as ``vortex equations'', describing the fractional quantum Hall effect in 2-dimensional double-layered electron systems. Via variational methods, we prove the existence and uniqueness of multiple vortices over a doubly periodic domain and the full plane. In the doubly periodic situation, explicit sufficient and necessary conditions are obtained that relate the size of the domain and the vortex numbers. For the full plane case, existence is established for all finite-energy solutions and exponential decay estimates are proved. Quantization phenomena of the magnetic flux are found in both cases.
\end{abstract}

%%% ----------------------------------------------------------------------
\maketitle
%%% ----------------------------------------------------------------------
\section{Introduction}
In recent years, there has been an increasing interest in the study of field theoretic models governed by Chern-Simons theory \cite{JHan,Han,Lieb,CSLin,Nam,Spruck1, Spruck2,Spruck3, Spruck4, Tarantello1, Tarantello2, Tarantello3, Wang, Yang1, Yang2, Yang3, Yang4}. These models generally lead to very complicated systems of non-linear equations, which have presented new challenges to mathematical analysts. Particularly, in condensed matter physics, Chern-Simons theory can be used to describe the fractional quantum Hall effect (FQHE) \cite{Cha, Frolich1, Frolich2, Frolich3, Girvin1, Girvin2, Ichinose1, Ichinose2,Jackiw, Jain, Klitzing, Kohmoto, McDonald, Stone, Thoules1, Thoules2}. Ichinose and Sekiguchi \cite{Ichinose2} established an effective theory for topological excitations in the generic $(m,m,n)$ Halperin state in Chern-Simons gauge field theory, which describes the FQHE in double-layer 2-dimensional electron systems. Under the radially symmetric ansatz, Ichinose and Sekiguchi \cite{Ichinose2}, obtained a system of Toda-type equations, limited to single soliton configurations, and studied the numerical and qualitative behavior of soliton solutions.

In section 2, from the Lagrangian of the model, we get a coupled gauged Schr\"{o}dinger equations governing the bosonized electron fields in the upper and lower layers, as well as, a set of constraint equations arising from the electromagnetic vector potential. In the static case and by ignoring long-range inter-layer and intra-layer Coulomb repulsions, we reduce the afore-mentioned equations to a first-order system via a ``first integral''. The reduced system is of the BPS type (named after the seminal works of Bogomol'ny \cite{Bogo} and Prasad-Sommerfield \cite{Prasad}). We give an energy lower bound in terms of the total flux and show that all finite-energy solutions over the full plane are topological in our context. The BPS system is then transformed into a coupled non-linear elliptic system or ``vortex equations'', which is a general form of the Toda-type equations of Ichinose and Sekiguchi \cite{Ichinose2}. We present three sharp theorems establishing the existence, uniqueness, and exponential decay of the vortex solutions, in addition to, quantization of the magnetic flux.

In section 3, we consider vortex solutions over a doubly periodic domain. Via a variational principle and a direct minimization problem, we use a weak compactness argument to prove the existence of vortices. We establish necessary and sufficient conditions, as well as, the uniqueness of the solutions. By a simple integration, we get quantized magnetic flux formulas. The necessary and sufficient conditions for existence give limits on the vortex numbers in terms of the domain size.

In section 4 and 5, over the full plane, we establish the existence, uniqueness, and exponential decay estimates of all finite-energy solutions. Using a Choleski transformation, we find a variational structure of the elliptic system. Through the weakly lower semi-continuity, Gateaux differentiability, and strict convexity of an energy functional, we prove the existence and uniqueness of vortex solutions. Additionally, we exploit the decay estimates to exhibit quantized magnetic flux formulas. 

\section{BPS and Vortex Equations}
The Lagrangian describing the FQHE in double-layer 2-dimensional electron systems \cite{Ichinose2} is composed of two terms, the matter term $\mathcal{L}_{\phi}$ and the Chern-Simons term $\mathcal{L}_{CS}$, 
\begin{equation}
\mathcal{L}=\mathcal{L}_{\phi}+\mathcal{L}_{CS},\label{Lagrangian}
\end{equation}
where 
\begin{align}
\mathcal{L}_{\phi}&=i\overline{\psi}_{\uparrow}(\partial_0-ia_0^+-ia_0^-)\psi_{\uparrow}+i\overline{\psi}_{\downarrow}(\partial_0-ia_0^++ia_0^-)\psi_{\downarrow}\\
&-\dfrac{1}{2M}\sum\limits_{\sigma=\uparrow,\downarrow}{|D_j^{\sigma}\psi_{\sigma}|^2}-V(\psi_{\uparrow},\psi_{\downarrow}),\nonumber\\
\mathcal{L}_{CS}&=\mathcal{L}_{CS}(a^+_{\mu})+\mathcal{L}_{CS}(a^-_{\mu})\nonumber\\
&=-\dfrac{1}{4}\epsilon_{\mu\nu\lambda}\bigg(\dfrac{1}{p}a^+_{\mu}\partial_{\nu}a^+_{\lambda}+\dfrac{1}{q}a^-_{\mu}\partial_{\nu}a^-_{\lambda}\bigg).
\end{align}
There are two bosonized electron fields, which represent electrons in the upper and lower layers, denoted by $\psi_{\sigma}$ ($\sigma=1,2$ or $\uparrow, \downarrow$), respectively. $M$ is the mass of the electrons and $p$ and $q$ are nonzero real numbers. $a^+_{\mu}$ and $a^-_{\mu}$ are scalar potential fields corresponding to the $U(1)\otimes U(1)$ local symmetry. $\mu$, $\nu$, and $\lambda$ take on the values $0,1,2$ and $\epsilon_{\mu\nu\lambda}$ is the antisymmetric metric tensor. Denote $\epsilon_{jk}=\epsilon_{0jk}$. The gauge-covariant derivatives are defined as 
\begin{equation}
D^{\uparrow\downarrow}_j=\partial_j-ia^+_j\mp ia_j^-+ieA_j,\qquad j=1,2,
\end{equation} 
where the external magnetic field, $B$, is directed along the $z$-axis and in the symmetric Coulomb gauge $A_j=-\dfrac{B}{2}\epsilon_{jk}x_k$.  We denote the temporal derivatives by
\begin{equation}
D^{\uparrow\downarrow}_0=\partial_0-ia^+_0\mp ia_0^-.
\end{equation}
The term $V(\psi_{\uparrow},\psi_{\downarrow})$ represents the interaction between electrons like the Coulomb repulsion and short-range four-body interaction. $V(\psi_{\uparrow},\psi_{\downarrow})$ is given by  
\begin{equation}
V(\psi_{\uparrow},\psi_{\downarrow})=\frac{p}{M}\left(\bar{\psi}_{\uparrow}\psi_{\uparrow}+\bar{\psi}_{\downarrow}\psi_{\downarrow}\right)^2+\frac{q}{M}\left(\bar{\psi}_{\uparrow}\psi_{\uparrow}-\bar{\psi}_{\downarrow}\psi_{\downarrow}\right)^2+W(\psi_{\uparrow},\psi_{\downarrow}),
\end{equation}
and more specifically represents the potential between bosonized electrons. $W(\psi_{\uparrow},\psi_{\downarrow})$ is the long-range inter-layer and intra-layer Coulomb repulsions. The corresponding Euler-Lagrange equations of the Lagrangian density \eqref{Lagrangian} are
\begin{subequations}
\begin{align}
iD_0^{\uparrow}\psi_{\uparrow}&=-\dfrac{1}{2M}(D_1^{\uparrow}D_1^{\uparrow}+D_2^{\uparrow}D_2^{\uparrow})\psi_{\uparrow}\label{gaugeSchro1}\\
&+\dfrac{2p}{M}(\bar{\psi}_{\uparrow}\psi_{\uparrow}+\bar{\psi}_{\downarrow}\psi_{\downarrow})\psi_{\uparrow}+\dfrac{2q}{M}(\bar{\psi}_{\uparrow}\psi_{\uparrow}-\bar{\psi}_{\downarrow}\psi_{\downarrow})\psi_{\uparrow}\nonumber\\
&+\frac{e^2}{4\pi\epsilon_0}\int{\left\lbrace\frac{2\psi_{\uparrow}(x)\bar{\psi}_{\uparrow}(x')\psi_{\uparrow}(x')}{|x-x'|}+\frac{\psi_{\uparrow}(x')\bar{\psi}_{\downarrow}(x)\psi_{\downarrow}(x)}{\sqrt{d^2+|x-x'|^2}}\right\rbrace}dx'\nonumber\\
iD_0^{\downarrow}\psi_{\downarrow}&=-\dfrac{1}{2M}(D_1^{\downarrow}D_1^{\downarrow}+D_2^{\downarrow}D_2^{\downarrow})\psi_{\downarrow}\label{gaugeSchro2}\\
&+\dfrac{2p}{M}(\bar{\psi}_{\uparrow}\psi+\bar{\psi}_{\downarrow}\psi_{\downarrow})\psi_{\downarrow}-\dfrac{2q}{M}(\bar{\psi}_{\uparrow}\psi_{\uparrow}-\bar{\psi}_{\downarrow}\psi_{\downarrow})\psi_{\downarrow}\nonumber\\
&+\frac{e^2}{4\pi\epsilon_0}\int{\left\lbrace\frac{2\psi_{\downarrow}(x)\bar{\psi}_{\downarrow}(x')\psi_{\downarrow}(x')}{|x-x'|}+\frac{\bar{\psi}_{\uparrow}(x)\psi_{\uparrow}(x)\psi_{\downarrow}(x')}{\sqrt{d^2+|x-x'|^2}}\right\rbrace}dx'\nonumber\\
F_{12}&=2p\left(\bar{\psi}_{\uparrow}\psi_{\uparrow}+\bar{\psi}_{\downarrow}\psi_{\downarrow}\right)\label{CSconstraint1}\\
\tilde{F}_{12}&=2q\left(\bar{\psi}_{\uparrow}\psi_{\uparrow}-\bar{\psi}_{\downarrow}\psi_{\downarrow}\right)\label{CSconstraint2}\\
F_{20}&=i\frac{p}{M}\bigg[\left(\psi_{\uparrow}\overline{D_1^{\uparrow}\psi_{\uparrow}}-\bar{\psi}_{\uparrow}D_1^{\uparrow}\psi_{\uparrow}\right)+\left(\psi_{\downarrow}\overline{D_1^{\downarrow}\psi_{\downarrow}}-\bar{\psi}_{\downarrow}D_1^{\downarrow}\psi_{\downarrow}\right)\bigg]\label{constraint1}\\
\tilde{F}_{20}&=i\frac{q}{M}\bigg[\left(\psi_{\uparrow}\overline{D_1^{\uparrow}\psi_{\uparrow}}-\bar{\psi}_{\uparrow}D_1^{\uparrow}\psi_{\uparrow}\right)-\left(\psi_{\downarrow}\overline{D_1^{\downarrow}\psi_{\downarrow}}-\bar{\psi}_{\downarrow}D_1^{\downarrow}\psi_{\downarrow}\right)\bigg]\label{constraint2}\\
F_{01}&=i\frac{p}{M}\bigg[\left(\psi_{\uparrow}\overline{D_2^{\uparrow}\psi_{\uparrow}}-\bar{\psi}_{\uparrow}D_2^{\uparrow}\psi_{\uparrow}\right)+\left(\psi_{\downarrow}\overline{D_2^{\downarrow}\psi_{\downarrow}}-\bar{\psi}_{\downarrow}D_2^{\downarrow}\psi_{\downarrow}\right)\bigg]\label{constraint3}\\
\tilde{F}_{01}&=i\frac{q}{M}\bigg[\left(\psi_{\uparrow}\overline{D_2^{\uparrow}\psi_{\uparrow}}-\bar{\psi}_{\uparrow}D_2^{\uparrow}\psi_{\uparrow}\right)-\left(\psi_{\downarrow}\overline{D_2^{\downarrow}\psi_{\downarrow}}-\bar{\psi}_{\downarrow}D_2^{\downarrow}\psi_{\downarrow}\right)\bigg],\label{constraint4}
\end{align}
\end{subequations}
where $F_{\mu\nu}=\partial_{\mu}a^+_{\nu}-\partial_{\nu}a^+_{\mu}$ and $\tilde{F}_{\mu\nu}=\partial_{\mu}a^-_{\nu}-\partial_{\nu}a^-_{\mu}$. We recognize equations \eqref{gaugeSchro1} and \eqref{gaugeSchro2} as a coupled gauged Schr\"{o}dinger equations governing the bosonized electron fields in the upper and lower layers, respectively. Equations \eqref{CSconstraint1} and \eqref{CSconstraint2} are the Chern-Simons constraints. As a result of not treating the electromagnetic potentials $a^+=(a^+_{\mu})$ and $a^-=(a^-_{\mu})$ as background fields, we get the constraint equations \eqref{constraint1}-\eqref{constraint4}.

In this paper, we will assume that the long-range inter-layer and intra-layer Coulomb repulsions are negligible and focus on the static case of the Euler-Lagrange equations. With these assumptions, and letting $B_{\mu\nu}=\partial_{\mu}b_{\nu}-\partial_{\nu}b_{\mu}-eB$ and $\tilde{B}_{\mu\nu}=\partial_{\mu}\tilde{b}_{\nu}-\partial_{\nu}\tilde{b}_{\mu}-eB$ represent the rate of change of the magnetic fields, where $b_{\mu}=a_{\mu}^++a_{\mu}^-$ and $\tilde{b}_{\mu}=a_{\mu}^+-a_{\mu}^-$, the system \eqref{gaugeSchro1}-\eqref{constraint4} is equivalent to
\begin{subequations}
\begin{align}
b_0\psi_{\uparrow}&=-\dfrac{1}{2M}D_j^{\uparrow}D_j^{\uparrow}\psi_{\uparrow}+\dfrac{2}{M}(p+q)|\psi_{\uparrow}|^2\psi_{\uparrow}+\dfrac{2}{M}(p-q)|\psi_{\downarrow}|^2\psi_{\uparrow}\label{StaticSchro1}\\
\tilde{b}_0\psi_{\downarrow}&=-\dfrac{1}{2M}D_j^{\downarrow}D_j^{\downarrow}\psi_{\downarrow}+\dfrac{2}{M}(p-q)|\psi_{\uparrow}|^2\psi_{\downarrow}+\dfrac{2}{M}(p+q)|\psi_{\downarrow}|^2\psi_{\downarrow}\label{StaticSchro2}\\
B_{12}&=2(p+q)|\psi_{\uparrow}|^2+2(p-q)|\psi_{\downarrow}|^2-eB\label{StaticCSconstraint1}\\
\tilde{B}_{12}&=2(p-q)|\psi_{\uparrow}|^2+2(p+q)|\psi_{\downarrow}|^2-eB\label{StaticCSconstraint2}\\
\partial_1 b_0&=\partial_1\tilde{b}_0=-\frac{i}{M}(p-\epsilon_{\sigma}q)\left(\psi_{\sigma}\overline{D_2^{\sigma}\psi_{\sigma}}-\bar{\psi}_{\sigma}D_2^{\sigma}\psi_{\sigma}\right)\label{staticConstraint1}\\
\partial_2 b_0&=\dfrac{i}{M}(p+\epsilon_{\sigma}q)\left(\psi_{\sigma}\overline{D_1^{\sigma}\psi_{\sigma}}-\bar{\psi}_{\sigma}D_1^{\sigma}\psi_{\sigma}\right)\label{staticConstraint2}\\
\partial_2 \tilde{b}_0&=\dfrac{i}{M}(p-\epsilon_{\sigma}q)\left(\psi_{\sigma}\overline{D_1^{\sigma}\psi_{\sigma}}-\bar{\psi}_{\sigma}D_1^{\sigma}\psi_{\sigma}\right).\label{staticConstraint3}
\end{align}
\end{subequations}
Above, we use the summation convention on the indexes $j$ and $\sigma$ and define $\epsilon_{\uparrow}=1$, $\epsilon_{\downarrow}=-1$. The system \eqref{StaticCSconstraint1}-\eqref{staticConstraint3} is similar to the non-relativistic Chern-Simons theory due to Jackiw and Pi \cite{Jackiw} and discussed by Yang \cite{Yang4}. Recall that for any complex-valued functions $\psi$ and $\eta$, we have the identity 
\begin{equation}
\partial_{\mu}(\psi\bar{\eta})=\psi\overline{D_{\mu}\eta}+(D_{\mu}\psi)\bar{\eta}.
\end{equation}
Using this identity we get
\begin{align}
\partial_j b_0&=\dfrac{1}{M}(p-\epsilon_{jk}q)\partial_j|\psi_{\uparrow}|^2-\dfrac{p}{M}\psi_{\uparrow}\left(\overline{D_j^{\uparrow}\psi_{\uparrow}-i\epsilon_{jk}D_k^{\uparrow}\psi_{\uparrow}}\right)\label{identity1}\\
&+\epsilon_{jk}\dfrac{q}{M}\bar{\psi}_{\uparrow}\left(D_j^{\uparrow}\psi_{\uparrow}+i\epsilon_{jk}D_k^{\uparrow}\psi_{\uparrow}\right)\nonumber\\
&+\dfrac{1}{M}(p+\epsilon_{jk}q)\partial_j|\psi_{\downarrow}|^2\nonumber-\dfrac{p}{M}\psi_{\downarrow}\left(\overline{D_j^{\downarrow}\psi_{\downarrow}-i\epsilon_{jk}D_k^{\downarrow}\psi_{\downarrow}}\right)\\
&+\epsilon_{jk}\dfrac{q}{M}\bar{\psi}_{\downarrow}\left(D_j^{\downarrow}\psi_{\downarrow}+i\epsilon_{jk}D_k^{\downarrow}\psi_{\downarrow}\right)\nonumber
\end{align}
and a similar equation for $\partial_j \tilde{b}_0$. Introduce the operators $D^{\uparrow\downarrow}_{\pm}$ as follows,
\begin{equation}
D^{\uparrow}_{\pm}=D_1^{\uparrow}\pm iD_2^{\uparrow}
\text{ and }
D^{\downarrow}_{\pm}=D_1^{\downarrow}\pm iD_2^{\downarrow}.\label{operators}
\end{equation}
Using \eqref{operators} in \eqref{StaticSchro1} and \eqref{StaticSchro2} we get
\begin{subequations}
\begin{align}
b_0\psi_{\uparrow}&=-\dfrac{1}{2M}\left(D^{\uparrow}_{+}D^{\uparrow}_{-}+B_{12}-eB\right)\psi_{\uparrow} \\
&+\dfrac{2}{M}(p+q)|\psi_{\uparrow}|^2\psi_{\uparrow}+\dfrac{2}{M}(p-q)|\psi_{\downarrow}|^2\psi_{\uparrow}\nonumber,\\
\tilde{b}_0\psi_{\downarrow}&=-\dfrac{1}{2M}\left(D^{\downarrow}_{+}D^{\downarrow}_{-}+\tilde{B}_{12}-eB\right)\psi_{\downarrow}\label{identity2} \\
&+\dfrac{2}{M}(p-q)|\psi_{\uparrow}|^2\psi_{\downarrow}+\dfrac{2}{M}(p+q)|\psi_{\downarrow}|^2\psi_{\downarrow}\nonumber.
\end{align}
\end{subequations}
As a consequence of identities \eqref{identity1} to \eqref{identity2}, we obtain a ``first integral" of the system \eqref{StaticCSconstraint1}-\eqref{staticConstraint3} of the BPS type,
\begin{subequations}
\begin{align}
D_{-}^{\uparrow}\psi_{\uparrow}&=(D_1^{\uparrow}-iD_2^{\uparrow})\psi_{\uparrow}=0\label{selfdual1}\\
D_{-}^{\downarrow}\psi_{\downarrow}&=(D_1^{\downarrow}-iD_2^{\downarrow})\psi_{\downarrow}=0\label{selfdual2}\\
B_{12}&=2(p+q)|\psi_{\uparrow}|^2+2(p-q)|\psi_{\downarrow}|^2-eB\label{magnetic1}\\
\tilde{B}_{12}&=2(p-q)|\psi_{\uparrow}|^2+2(p+q)|\psi_{\downarrow}|^2-eB\label{magnetic2}\\
b_0&=\dfrac{1}{M}(p+q)|\psi_{\uparrow}|^2+\dfrac{1}{M}(p-q)|\psi_{\downarrow}|^2+\dfrac{eB}{M}\label{field1}\\
\tilde{b}_0&=\dfrac{1}{M}(p- q)|\psi_{\uparrow}|^2+\dfrac{1}{M}(p+ q)|\psi_{\downarrow}|^2+\dfrac{eB}{M}.\label{field2}
\end{align}
\end{subequations}
Consider the total energy \cite{Ichinose2}, 
\begin{align}
E=\dfrac{1}{2M}\int{\left\{\sum_{\sigma=\uparrow,\downarrow}|(D_1^{\sigma}-iD_2^{\sigma})\psi_{\sigma}|^2+eB\left(|\psi_{\uparrow}|^2+|\psi_{\downarrow}|^2\right)-eB\left(|\psi_{\uparrow,0}|^2+|\psi_{\downarrow,0}|^2\right)\right\}}dx,\label{energy}
\end{align}
where $\psi_{\uparrow,0}=\sqrt{\bar{\rho}/2}$ and $\psi_{\downarrow,0}=\sqrt{\bar{\rho}/2}$ are the field configurations for the ground state of the fractional quantum Hall effect and $\overline{\rho}$ is the average electron density. Additionally, the ground state for the fractional quantum Hall effect requires the filling factor $\nu$ to satisfy 
\begin{equation}
\nu=\frac{2\pi\overline{\rho}}{eB}=\frac{\pi}{p}.\label{fillingfactor}
\end{equation}
As in \cite{Ichinose2} and using equations \eqref{magnetic1} and \eqref{magnetic2}, we define the total Chern-Simons flux by
\begin{align}
\Phi_{CS}=\dfrac{1}{2}\int\left( B_{12}+\tilde{B}_{12}\right)dx.
\end{align}
Hence, we may write the following energy lower bound
\begin{align}
E\geq\dfrac{eB}{4Mp}\Phi_{CS},
\end{align}
and note that solutions to the ``self-dual" equations \eqref{selfdual1} and \eqref{selfdual2} correspond to lowest energy configurations.
 
We are interested in finite-energy solutions of the system \eqref{selfdual1}-\eqref{field2}. Using \eqref{selfdual1} and \eqref{selfdual2}, the finite-energy condition reduces to
\begin{align}
E=\dfrac{eB}{2M}\int\left(|\psi_{\uparrow}|^2+|\psi_{\downarrow}|^2-\left(|\psi_{\uparrow,0}|^2+|\psi_{\downarrow,0}|^2\right)\right) dx<\infty.\label{finiteEnergy}
\end{align}
In particular, over the full plane $\mathbb{R}^2$ and to guarantee that the amplitude of the electron fields are equal to that of the ground state, we see that the finite-energy condition imposes the boundary condition:
\begin{align}
|\psi_{\uparrow}|^2\rightarrow |\psi_{\uparrow,0}|^2=\dfrac{\bar{\rho}}{2}\qquad\text{and} \qquad
|\psi_{\downarrow}|^2\rightarrow |\psi_{\downarrow,0}|^2=\dfrac{\bar{\rho}}{2}\qquad\text{as}\qquad |x|\rightarrow\infty.\label{topBC}
\end{align}
Solutions satisfying \eqref{topBC} are called topological. Therefore, in our context, all finite-energy solutions over the full plane are topological. 

To arrive at the coupled non-linear elliptic system of interest (i.e, ``vortex equations'') we first introduce the complexified variables $\partial=\partial_1+i\partial_2$, $\beta=b_1+ib_2$, and $A=A_1+iA_2$. Away from the zeros of the field $\psi_{\uparrow}$, from the self-dual equation \eqref{selfdual1}, we obtain the system
\begin{subequations}
\begin{align}
\partial\overline{\partial}\ln|\psi_{\uparrow}|&=i\partial\overline{\beta}-ie\partial\overline{A}\label{eq1}\\
\overline{\partial}\partial\ln|\overline{\psi}_{\uparrow}|&=-i\overline{\partial}\beta+ie\overline{\partial}A.\label{eq2}
\end{align}
\end{subequations}
Summing \eqref{eq1} and \eqref{eq2}, and using $\Delta=\partial\overline{\partial}=\overline{\partial}\partial$, we get
\begin{equation}
\Delta\ln|\psi_{\uparrow}|^2=i(\partial\overline{\beta}-\overline{\partial}\beta)-ie(\partial\overline{A}-\overline{\partial}A).
\end{equation}
Let us note that 
\begin{equation}
\partial\overline{\beta}=\mbox{div}(\beta)-i(\partial_1b_2-\partial_2b_1) \quad\text{and}\quad \partial\overline{A}=-iB.
\end{equation}
Hence,
\begin{equation}
\Delta\ln|\psi_{\uparrow}|^2=4(p+q)|\psi_{\uparrow}|^2+4(p-q)|\psi_{\downarrow}|^2-4p\overline{\rho}.
\end{equation}
In a similar manner, we may use the self-dual equation \eqref{selfdual2} to obtain 
\begin{equation}
\Delta\ln|\psi_{\downarrow}|^2=4(p-q)|\psi_{\uparrow}|^2+4(p+q)|\psi_{\downarrow}|^2-4p\overline{\rho}.
\end{equation}
Let us denote the set of zeros of the fields  $\psi_{\uparrow}$ and $\psi_{\downarrow}$ by $Z_{\psi_{\uparrow}}=\{p_1,\ldots,p_{N_1}\}$ and $Z_{\psi_{\downarrow}}=\{q_1,\dots,q_{N_2}\}$, respectively. Note that the zeros of the fields $\psi_{\uparrow}$ and $\psi_{\downarrow}$ are discrete and of integer multiplicities. Define the functions $u_1$ and $u_2$ by 
\begin{equation}
u_1=\ln|\psi_{\uparrow}|^2-\ln|\bar{\rho}|\quad\text{and}\quad u_2=\ln|\psi_{\downarrow}|^2-\ln|\bar{\rho}|.
\end{equation}
Then, using the transformation $x\mapsto\sqrt{p\overline{\rho}}x$, we arrive at the vortex equations,
\begin{subequations}
\begin{align}
\Delta u_1 &=  4\left(k_{11}e^{u_1}+k_{12}e^{u_2}-1\right)+4\pi\sum\limits_{j=1}^{N_1}{\delta_{p_j}(x)}\label{elliptic1}\\
\Delta u_2 &=4\left(k_{21}e^{u_1}+k_{22}e^{u_2}-1\right)+4\pi\sum\limits_{j=1}^{N_2}{\delta_{q_j}(x)}\label{elliptic2},
\end{align}
\end{subequations}
defined for all $x\in\mathbb{R}^2$, where $K=(k_{ij})$ is the matrix
\begin{equation}
K=\frac{1}{p}\left(
\begin{array}{cc}
p+q & p-q\\
p-q & p+q
\end{array}\right),\label{matrixK}
\end{equation}
and $\delta_{P}(x)$ is the Dirac distribution concentrated at the point $P$. The positive integers $N_1$ and $N_2$ are called the vortex numbers. In this paper, we are only interested in the coupled system, for which $p$ not equal to $q$.

We now state our main existence, uniqueness, asymptotic behaviour, and quantized magnetic flux theorems. In what follows, the notation $|\Omega|$ and $|K|$, denotes the size of the domain and the determinant of the matrix $K$, respectively. 

\begin{thm}\label{doublyPeriodic} Consider the coupled non-linear elliptic system \eqref{elliptic1}-\eqref{elliptic2}, over a doubly periodic domain $\Omega$. Let the matrix $K$, given by \eqref{matrixK}, be positive definite. Then a unique solution exists if and only if the condition 
\begin{equation}
|\Omega|>\max\bigg\{\dfrac{2\pi}{|K|}\left(k_{22}N_1-k_{12}N_2\right),\dfrac{2\pi}{|K|}\left(k_{11}N_2-k_{21}N_1\right)\bigg\}
\end{equation}
is satisfied.
\end{thm}

\begin{thm}\label{fullPlane} Consider the coupled non-linear elliptic system \eqref{elliptic1}-\eqref{elliptic2} over the full plane $\mathbb{R}^2$ and satisfying the boundary condition
\begin{equation}
u_1,u_2\rightarrow -\ln(2)\qquad\text{as}\qquad |x|\rightarrow\infty.\label{topobound}
\end{equation} Let the matrix $K$, given by \eqref{matrixK}, be positive definite. There exists a unique solution to the system, satisfying the boundary condition exponentially fast. More precisely, we have the following exponential decay estimate, 
\begin{equation}
(u_1+\ln 2)^2+(u_2+\ln 2)^2\leq C_{\epsilon}e^{-(1-\epsilon)\sqrt{\lambda_0}|x|}\label{2.46}
\end{equation} 
when $|x|$ is sufficiently large, $\epsilon\in(0,1)$ is arbitrary, $C_{\epsilon}>0$ is a constant depending on $\epsilon$, and 
$\lambda_0= 4 \min \left\{ 2,\frac{2q}{p}\right\}$. Additionally, 
\begin{equation}
|\nabla u_1|^2+|\nabla u_2|^2\leq C_{\delta}e^{-(1-\delta)\sqrt{\lambda_0}|x|},\label{2.47}
\end{equation}
when $|x|$ is sufficiently large, $\delta\in(0,1)$ is arbitrary, and $C_{\delta}>0$ is a constant depending on $\delta$.
\end{thm}

\begin{thm}\label{quantizedInt}
In both the doubly periodic and the full plane cases, considered in Theorem \ref{doublyPeriodic} and Theorem \ref{fullPlane}, there hold the quantized magnetic flux integrals
\begin{equation}
\int{B_{12}}dx=-2\pi pN_1\qquad\text{and}\qquad
\int{\tilde{B}_{12}}dx=-2\pi pN_2,
\end{equation} 
where the integration is evaluated either over the doubly periodic domain $\Omega$ or the full plane $\R^2$.
\end{thm}

\section{Solution Over A Doubly Periodic Domain}
In this section, we prove Theorem \ref{doublyPeriodic}. We first establish necessary conditions for the existence of a solution and derive the quantized magnetic flux integrals of Theorem \ref{quantizedInt}, over a doubly periodic domain. Through a Choleski transformation, as in \cite{Yang1,Yang4}, we find a variational principle. We establish a coercivity condition and via a direct minimization approach, prove the existence of a solution to the system \eqref{elliptic1}-\eqref{elliptic2} over a doubly periodic domain. 

By a doubly periodic domain, $\Omega$, we mean a periodic lattice cell with a ``gauge-periodic" boundary \cite{Yang4}. In other words, our solutions are periodic over a cell domain modulo gauge transformations as introduced by 't Hooft \cite{Hooft}. In this concrete situation, we can identify $\Omega$ with the 2-torus $\Omega=\R^2/\Omega$.

There are functions $u_0^{'}:\Omega\rightarrow\mathbb{R}$ and $u_0^{''}:\Omega\rightarrow\mathbb{R}$, whose existence and uniqueness (up to an additive constant) are guaranteed by Aubin in \cite{Aubin}, satisfying
\begin{equation}
\Delta u_0^{'}=-\dfrac{4\pi N_1}{|\Omega|}+4\pi\sum_{j=1}^{N_1}\delta_{p_j}(x)\quad\text{and}\quad
\Delta u_0^{''}=-\dfrac{4\pi N_2}{|\Omega|}+4\pi\sum_{j=1}^{N_2}\delta_{q_j}(x).
\end{equation}
Let $u_1,u_2:\Omega\rightarrow\mathbb{R}$ be functions satisfying the system \eqref{elliptic1}-\eqref{elliptic2}. Define $v_1=u_1-u_0^{'}$ and $v_2=u_2-u_0^{''}$ on $\Omega$. Then, 
\begin{subequations}
\begin{align}
\Delta v_1 &=4(k_{11}e^{u_0^{'}+v_1}+k_{12}e^{u_0^{''}+v_2}-1)+\dfrac{4\pi N_1}{|\Omega|}\label{ellpv1}\\
\Delta v_2 &=4(k_{21}e^{u_0^{'}+v_1}+k_{22}e^{u_0^{''}+v_2}-1)+\dfrac{4\pi N_2}{|\Omega|}\label{ellpv2}.
\end{align}
\end{subequations}
Integrating \eqref{ellpv1}-\eqref{ellpv2}, we obtain the following linear system of equations in the unknowns $\int_{\Omega}{e^{u_0^{'}+v_1}}dx$ and $\int_{\Omega}{e^{u_0^{''}+v_2}}dx$,
\begin{subequations}
\begin{align}
k_{11}\int_{\Omega}{e^{u_0^{'}+v_1}}dx+k_{12}\int_{\Omega}{e^{u_0^{''}+v_2}}dx&=|\Omega|-\pi N_1\\
k_{21}\int_{\Omega}{e^{u_0^{'}+v_1}}dx+k_{22}\int_{\Omega}{e^{u_0^{''}+v_2}}dx&=|\Omega|-\pi N_2.
\end{align}
\end{subequations}
Solving for $\int_{\Omega}{e^{u_0^{'}+v_1}}dx$ and $\int_{\Omega}{e^{u_0^{''}+v_2}}dx$ we obtain the necessary conditions
\begin{subequations}
\begin{align}
\int_{\Omega}{e^{u_0^{'}+v_1}}dx&=\frac{|\Omega|}{2}-(k_{22}N_1-k_{12}N_2)\frac{\pi}{|K|}\equiv \eta_1>0\label{ineq1}\\
\int_{\Omega}{e^{u_0^{''}+v_2}}dx&=\frac{|\Omega|}{2}-(k_{11}N_2-k_{21}N_1)\frac{\pi}{|K|}\equiv \eta_2>0.\label{ineq2}
\end{align}
\end{subequations}
In terms of the size of the domain, inequalities \eqref{ineq1} and \eqref{ineq2} give
\begin{equation}
|\Omega|>\max\bigg\{\dfrac{2\pi}{|K|}\left(k_{22}N_1-k_{12}N_2\right),\dfrac{2\pi}{|K|}\left(k_{11}N_2-k_{21}N_1\right)\bigg\},
\end{equation}
and the necessity condition of Theorem \ref{doublyPeriodic} is established. Note that the vortex numbers  $N_1$ and $N_2$ are constrained by the size of the domain. 

Integrating equations \eqref{ellpv1} and \eqref{ellpv2} over $\Omega$, and expressing the results in terms of $|\psi_{\sigma}|^2$, we get
\begin{subequations}
\begin{align}
\int_{\Omega}{(2(p+q)|\psi_{\uparrow}|^2+2(p-q)|\psi_{\downarrow}|^2-2p)}dx&=-2\pi p N_1\\
\int_{\Omega}{(2(p-q)|\psi_{\uparrow}|^2+2(p+q)|\psi_{\downarrow}|^2-2p)}dx&=-2\pi pN_2.
\end{align}
\end{subequations}
Therefore, from \eqref{magnetic1}, \eqref{magnetic2}, and \eqref{fillingfactor} we arrive at the quantized magnetic flux formulas of Theorem \ref{quantizedInt}, 
\begin{equation}
\int_{\Omega}{B_{12}}dx=-2\pi p N_1\qquad\text{and}\qquad
\int_{\Omega}{\tilde{B}_{12}}dx=-2\pi pN_2.
\end{equation}

In its current form, the system \eqref{ellpv1}-\eqref{ellpv2} does not have a simple variational structure. However, when the matrix $K$ is positive definite, a variational principle for the elliptic system can be found via a Choleski decomposition, i.e., there is a unique lower triangular matrix $L$ such that $K=LL^t$. To this end, consider the Choleski transformation
\begin{equation}
w_1=\frac{1}{\sqrt{|K|}}v_1\quad\text{and}\quad
w_2=\frac{1}{|K|}(k_{11}v_2-k_{21}v_1).\label{Choleski}
\end{equation}
The system \eqref{ellpv1}-\eqref{ellpv2} becomes
\begin{subequations}
\begin{align}
\Delta w_1 &= \frac{4k_{11}}{\sqrt{|K|}}e^{u_0^{'}+\sqrt{|K|}w_1}+\frac{4k_{12}}{\sqrt{|K|}}e^{u_0^{''}+(|K|w_2+k_{21}\sqrt{|K|}w_1)/k_{11}}-C_1\\
\Delta w_2 &=4 e^{u_0^{''}+(|K|w_2+k_{21}\sqrt{|K|}w_1)/k_{11}}-C_2,
\end{align}
\end{subequations}
where 
\begin{equation}
C_1=\dfrac{4}{\sqrt{|K|}}\bigg(1-\dfrac{\pi N_1}{|\Omega|}\bigg)\quad\text{and}\qquad
C_2=2-\dfrac{4\pi}{|\Omega||K|}\bigg(k_{11}N_2-k_{21}N_1\bigg).
\end{equation}
The corresponding functional $I:H^1_2\rightarrow \mathbb{R}$ is 
\begin{align}
I(w_1,w_2)&=\int_{\Omega}\bigg\{\dfrac{1}{2}|\nabla w_1|^2+\dfrac{1}{2}|\nabla w_2|^2+\frac{4k_{11}}{|K|}e^{u_0^{'}
+\sqrt{|K|}w_1}\label{functional}\\
&+\frac{4k_{11}}{|K|}e^{u_0^{''}+(|K|w_2+k_{21}\sqrt{|K|}w_1)/k_{11}}-C_1w_1-C_2 w_2\bigg\}dx.\nonumber
\end{align}

Let us show that the above functional $I$ satisfies a coercive lower bound. We use the notation $H^1_2=W^{1,2}(\R^2)\times W^{1,2}(\R^2)$, $H^1=W^{1,2}(\R^2)$, and $||\cdot||_{1,2}$ to denote the norm of $H^1_2$. 

Decompose $H^1$ as the direct sum of $\mathbb{R}$ and the set $\tilde{H}^1$, i.e.  $H^1=\mathbb{R}\oplus\tilde{H}^1$. The set $\tilde{H}^1$ is defined as the collection of all $\tilde{w}\in H^1$ such that $\int_{\Omega}{\tilde{w}}dx=0$. Let  $w_1=\dot{w}_1+\tilde{w}_1$ and $w_2=\dot{w}_2+\tilde{w}_2$, where $\dot{w}_1,\dot{w}_2\in\mathbb{R}$ and $\tilde{w}_1,\tilde{w}_2\in\tilde{H}^1(\Omega)$.
From the necessary conditions \eqref{ineq1} and \eqref{ineq2}, we get
\begin{subequations}
\begin{align}
\sqrt{|K|}\dot{w}_1&=\ln(\eta_1)-\ln\bigg(\int_{\Omega}{e^{u_0^{'}+\sqrt{|K|}\tilde{w}_1}}dx\bigg)\label{eq3}\\
(|K|\dot{w}_2+k_{21}\sqrt{|K|}\dot{w}_1)/k_{11}&=\ln(\eta_2)
-\ln\bigg(\int_{\Omega}{e^{u_0^{''}+(|K|\tilde{w}_2+k_{21}\sqrt{|K|}\tilde{w}_1)/k_{11}}}dx\bigg).
\end{align}
\end{subequations}
Applying the above decomposition to the functional \eqref{functional},
\begin{align*}
I(w_1,w_2)-&\dfrac{1}{2}\sum\limits_{i=1}^{2}{||\nabla \tilde{w}_i||^2_2}
=\dfrac{4k_{11}}{|K|}\eta_1\bigg(1-\ln(\eta_1)+\ln\bigg(\int_{\Omega}{e^{u_0^{'}+\sqrt{|K|}\tilde{w}_1}}dx\bigg)\bigg)\\
&+\dfrac{4k_{11}}{|K|}\eta_2\bigg(1-\ln(\eta_2)+\ln\bigg(\int_{\Omega}{e^{u_0^{''}+(|K|\tilde{w}_2+k_{21}\sqrt{|K|}\tilde{w}_1)/k_{11}}}dx\bigg)\bigg).\nonumber
\end{align*}
Note that $k_{11}>0$, since $|K|>0$. By Jensen's inequality we obtain 
\begin{subequations}
\begin{align}
\ln\bigg(\int_{\Omega}{e^{u_0^{'}+\sqrt{|K|}\tilde{w}_1}}dx\bigg)&\geq\dfrac{1}{|\Omega|} \int_{\Omega}{u_0'}dx+\ln|\Omega|,\\
\ln\bigg(\int_{\Omega}{e^{u_0''+(|K|\tilde{w}_2+k_{21}\sqrt{|K|}\tilde{w}_1)/k_{11}}}dx\bigg) &\geq\dfrac{1}{|\Omega|}\int_{\Omega}{u_0''}dx+\ln|\Omega|,
\end{align}
\end{subequations}
and the coercive lower bound
\begin{align}
I(w_1,w_2)&\geq \dfrac{1}{2}\sum\limits_{i=1}^{2}{||\nabla w_i||^2_2-\beta}\label{coercive1}
\end{align}
is attained, where $\beta$ is a positive constant independent of $w_1$ and $w_2$.

From the coercive lower bound \eqref{coercive1}, we conclude that the functional $I$ is bounded from below. So it makes sense to consider the direct minimization problem:
\begin{equation}
m\equiv \inf\bigg\{I(w_1,w_2)\bigg|w_1,w_2\in H^1\bigg\}.\label{minimization}
\end{equation}
Let $\{(w_1^k,w_2^k)\}$ be a minimizing sequence of \eqref{minimization}, i.e., choose functions $w_1^k$ and $w_2^k$ in $H^1$, where $k=1,2,3,\ldots$ so that 
\begin{equation*}
I(w_1^k,w_2^k)\rightarrow m \quad\text{ as }\quad k\rightarrow \infty \quad \text{ and }\quad
I(w_1^1,w_2^1)\geq I(w_1^2,w_2^2)\geq \dots\geq m.
\end{equation*}
From \eqref{coercive1} and the decomposition $w_i^k=\dot{w}_i^k+\tilde{w}_i^k$ where $i=1,2$, we may write
\begin{equation*}
I(w_1^k,w_2^k)\geq\dfrac{1}{2}\sum\limits_{i=1}^{2}{||\nabla \tilde{w}_i^k||^2_2}-\beta,
\end{equation*} 
and conclude that $\{\nabla\tilde{w}_i^k\}$ all belong in $L^2(\Omega)$. 

Recall the Poincar\'{e} inequality 
\begin{equation*}
\int_{\Omega}{f^2(x)}dx\leq C\int_{\Omega}{|\nabla f(x)|^2}dx, \quad f\in W^{1,2}(\Omega),\quad \int_{\Omega}{f(x)}dx=0,
\end{equation*} 
where $C>0$ is a suitable constant. By Poincar\'{e}'s inequality we obtain
\begin{align}
I(w_1^k,w_2^k)&\geq\dfrac{1}{2}\sum\limits_{i=1}^{2}{||\nabla \tilde{w}_i^k||^2_2}-\beta
\geq\sum\limits_{i=1}^{2}{\alpha_i|| \tilde{w}_i^k||^2_2}-\beta,
\end{align}
for some suitable positive constants $\alpha_i$'s, $i=1,2$, and conclude that the sequence $\{\tilde{w}_i^k\}$ all belong in $L^2(\Omega)$. Therefore, $\{(\tilde{w}_1^k,\tilde{w}_2^k)\}$ is bounded in $H^1_2$. Moreover, $\tilde{w}_1^{(\infty)}$ and $\tilde{w}_2^{(\infty)}$ belong in $\tilde{H}^1$, since 
\begin{equation}
\bigg|\int_{\Omega}{\tilde{w}_i^{(\infty)}}dx\bigg|\leq|\Omega|^{1/2}||\tilde{w}_i^{(\infty)}-\tilde{w}_i^k||_2\rightarrow 0 \quad \text{as}\quad k\rightarrow\infty.
\end{equation}
After all, we are in a reflexive space and without loss of generality, we may suppose that 
\begin{equation}
(\tilde{w}_1^k,\tilde{w}_2^k)\rightharpoonup (\tilde{w}_1^{(\infty)},\tilde{w}_2^{(\infty)})\in H^1_2 \qquad \text{ weakly as }\quad  k\rightarrow \infty.\label{weakconv1}
\end{equation}

Let us show that the sequences $\{\dot{w}^k_1\}$ and $\{\dot{w}^k_2\}$ of real numbers are also bounded. We will need the Trudinger-Moser inequality \cite{Aubin} of the form
\begin{equation}
\int_{\Omega}{e^{f(x)}}dx\leq C_1 e^{C_2 \int_{\Omega}{|\nabla f(x)|^2}dx},\quad f\in W^{1,2}(\Omega), \quad \int_{\Omega}{f(x)dx}=0, \label{Trud}
\end{equation} 
where $C_1$ and $C_2$ are positive constants.

From \eqref{eq3}, we may write
\begin{equation}
\sqrt{|K|}\dot{w}^k_1=\ln(\eta_1)-\ln\bigg(\int_{\Omega}{e^{u_0^{'}+\sqrt{|K|}\tilde{w}^k_1}}dx\bigg).\label{eq4}
\end{equation} 
Taking absolute value on both sides of \eqref{eq4} and enlarging we have
\begin{align}
|\sqrt{|K|}\dot{w}^k_1|
\leq&|\ln(\eta_1)|+\dfrac{1}{2}\bigg|\ln\bigg(\int_{\Omega}{e^{2u_0^{'}}}dx\bigg)\bigg|\label{eq5}\\
&+\dfrac{1}{2}\bigg|\ln\bigg(\int_{\Omega}{e^{2\sqrt{|K|}\tilde{w}^k_1}}dx\bigg)\bigg|.\nonumber
\end{align} 
Apply \eqref{Trud} to the last term in \eqref{eq5} to obtain
\begin{equation}
|\sqrt{|K|}\dot{w}^k_1|\leq|\ln(\eta_1)|+\dfrac{1}{2}\bigg|\ln\bigg(\int_{\Omega}{e^{2u_0^{'}}}dx\bigg)\bigg|+C_1C_2\sqrt{|K|}\bigg|\int_{\Omega}{\nabla\tilde{w}^k_1}dx\bigg|.
\end{equation}
Since the sequence $\{\tilde{w}_1^k\}$ is bounded in $H^1$, we conclude that the sequence $\{\dot{w}_1^k\}$ is bounded in $\mathbb{R}$. In a similar way, we may conclude that $\{\dot{w}_2^k\}$ is bounded in $\mathbb{R}$. Without loss of generality, we may assume that 
\begin{equation}
\dot{w}_i^k\rightarrow \dot{w}_i^{(\infty)}\in \mathbb{R}\qquad\text{ as } k\rightarrow\infty, \qquad i=1,2. \label{eq6}
\end{equation}
Define $w_i^{(\infty)}=\dot{w}_i^{(\infty)}+\tilde{w}_i^{(\infty)}$ for $i=1,2$. Therefore, from \eqref{weakconv1} and \eqref{eq6} we get that 
\begin{equation}
(w_1^k,w_2^k)\rightharpoonup (w_1^{(\infty)},w_2^{(\infty)})\in H^1_2 \qquad \text{weakly as }\quad  k\rightarrow \infty.
\end{equation}

To conclude that $(w_1^{(\infty)},w_2^{(\infty)})$ is the sought out solution of the minimization problem \eqref{minimization}, and hence a solution to the system \eqref{elliptic1}-\eqref{elliptic2}, we appeal to the weak lower semi-continuity of $I$. The uniqueness of the solution follows directly from the strict convexity of the functional $I$, which can be shown by a direct calculation of the corresponding Hessian matrix. Therefore, Theorem \ref{doublyPeriodic} is proved.

\section{Solution Over Full Plane $\R^2$}
In this section, via a variational principle, we prove the existence and uniqueness components of Theorem \ref{fullPlane}. Our approach follows the ideas by Jaffe and Taubes in \cite{Taubes} and Yang in \cite{Yang4}. In contrast to the doubly periodic case, the vortex numbers are not constrained by the domain size. 

In order to establish a variational principle, we transform the system \eqref{elliptic1}-\eqref{elliptic2}, satisfying the boundary conditions \eqref{topobound}, to an equivalent system, which can then be view as the Euler-Lagrange equations of a respective functional. 

Consider the background functions $u'_0,u''_0:\R^2\rightarrow\R$, depending on a real parameter $\mu>0$, defined by
\begin{align*}
u_0'(x)=-\sum\limits_{j=1}^{N_1}{\ln(1+\mu|x-p_j|^{-2})}\quad\text{and}\quad
u_0''(x)=-\sum\limits_{j=1}^{N_2}{\ln(1+\mu|x-q_j|^{-2})}.
\end{align*}
Note that $u'_0(x),u''_0(x)\leq 0$ for all $x\in\R^2$ and  

\begin{align*}
\Delta u_0'&=-4\sum\limits_{j=1}^{N_1}{\dfrac{\mu}{(\mu+|x-p_j|^2)^2}}+4\pi\sum\limits_{j=1}^{N_1}{\delta_{p_j}(x)},\\
\Delta u_0''&=-4\sum\limits_{j=1}^{N_1}{\dfrac{\mu}{(\mu+|x-q_j|^2)^2}}+4\pi\sum\limits_{j=1}^{N_2}{\delta_{q_j}(x)}.
\end{align*}

Let  
\begin{equation*}
g_0'(x)=4\sum\limits_{j=1}^{N_1}{\dfrac{\mu}{(\mu+|x-p_j|^2)^2}}\quad \text{ and }\quad g_0''(x)=4\sum\limits_{j=1}^{N_2}{\dfrac{\mu}{(\mu+|x-q_j|^2)^2}}.
\end{equation*}
We note that $u'_0$, $u''_0$, $g'_0$, and $g''_0$ all belong to $L^2(\R^2)$. Define  $v_1=u_1-u_0'$ and $v_2=u_2-u_0''$ on $\R^2$. Hence, the system \eqref{elliptic1}-\eqref{elliptic2} becomes 
\begin{subequations}
\begin{align}
\Delta v_1&=4\bigg(k_{11}e^{u_0'+v_1}+k_{12}e^{u_0''+v_2}-1\bigg)+g_0'\label{v1}\\
\Delta v_2&=4\bigg(k_{21}e^{u_0'+v_1}+k_{22}e^{u_0''+v_2}-1\bigg)+g_0'',\label{v2}
\end{align}
\end{subequations}
where $v_1(x)\rightarrow -\ln(2)$ and $v_2(x)\rightarrow -\ln(2)$ as $|x|\rightarrow\infty$. 

When $K$ is positive definite, we again use the transformation \eqref{Choleski}. 
Thus, the system \eqref{v1}-\eqref{v2} is transformed to
\begin{subequations}
\begin{align}
\Delta w_1 &= \frac{4K_{11}}{\sqrt{|K|}}\bigg(e^{u_0^{'}+\sqrt{|K|}w_1}-1\bigg)\\
&+\frac{4K_{12}}{\sqrt{|K|}}\bigg(e^{u_0^{''}+\frac{1}{K_{11}}(|K|w_2+K_{21}\sqrt{|K|}w_1)}-1\bigg)+h_1,\nonumber\\
\Delta w_2 &=4 \bigg(e^{u_0^{''}+\frac{1}{K_{11}}(|K|w_2+K_{21}\sqrt{|K|}w_1)}-1\bigg)+h_2,
\end{align}
\end{subequations}
where
\begin{equation}
h_1=\dfrac{1}{\sqrt{|K|}}(g_0'+4)\qquad\text{and}\qquad
h_2=2+\dfrac{1}{|K|}(K_{11}g_0''-K_{21}g_0').
\end{equation}
However, the functions $h_1$ and $h_2$ are not in $L^2(\R^2)$ and this is an undesired property for what follows. To correct this issue, we define $\tilde{v}_1=v_1+\ln(2)$ and $\tilde{v}_2=v_2+\ln(2)$. Using this new definition, the system \eqref{v1}-\eqref{v2} becomes  
\begin{subequations}
\begin{align}
\Delta \tilde{v}_1&=4\bigg(\dfrac{1}{2}k_{11}e^{u_0'+ \tilde{v}_1}+\dfrac{1}{2}k_{12}e^{u_0''+ \tilde{v}_2}-1\bigg)+g_0'\label{eq7}\\
\Delta \tilde{v}_2&=4\bigg(\dfrac{1}{2}k_{21}e^{u_0'+ \tilde{v}_1}+\dfrac{1}{2}k_{22}e^{u_0''+ \tilde{v}_2}-1\bigg)+g_0''\label{eq8},
\end{align}
\end{subequations}
where $\tilde{v}_1(x)\rightarrow 0$ and $\tilde{v}_2(x)\rightarrow 0$ as $|x|\rightarrow\infty$. 
With the use of \eqref{Choleski}, the system \eqref{eq7}-\eqref{eq8} is equivalent to
\begin{subequations}
\begin{align}
\Delta w_1 &= \frac{2k_{11}}{\sqrt{|K|}}\bigg(e^{u_0^{'}+\sqrt{|K|}w_1}-1\bigg)\label{w1}\\
&+\frac{2k_{12}}{\sqrt{|K|}}\bigg(e^{u_0^{''}+(|K|w_2+k_{21}\sqrt{|K|}w_1)/k_{11}}-1\bigg)+h_1\nonumber\\
\Delta w_2 &=2\bigg(e^{u_0^{''}+(|K|w_2+k_{21}\sqrt{|K|}w_1)/k_{11}}-1\bigg)+h_2,\label{w2}
\end{align}
\end{subequations}
where
\begin{equation}
h_1=\dfrac{1}{\sqrt{|K|}}g_0'\qquad\text{and}\qquad
h_2=\dfrac{1}{|K|}(K_{11}g_0''-K_{21}g_0').
\end{equation}
Moreover, $h_1,h_2\in L^2(\R^2)$, $w_1(x)\rightarrow 0$, and $w_2(x)\rightarrow 0$ as $|x|\rightarrow\infty$. 

The system \eqref{w1}-\eqref{w2} is well-defined and is the Euler-Lagrange equations of the functional $I:H^1_2\rightarrow\R$, given by 
\begin{align}
I(w_1,w_2)&=\int_{\mathbb{R}^2}{\bigg\{\frac{1}{2}|\nabla w_1|^2+\frac{1}{2}|\nabla w_2|^2+\frac{2k_{11}}{|K|}e^{u_0'}(e^{\sqrt{|K|}w_1}-1)\bigg\}}dx\label{functional2}\\
&+\int_{\mathbb{R}^2}{\bigg\{\frac{2k_{11}}{|K|}e^{u_0''}(e^{(k_{21}\sqrt{|K|}w_1+|K|w_2)/k_{11}}-1)\bigg\}}dx\nonumber\\
&+\int_{\mathbb{R}^2}{\bigg\{\bigg(h_1-\frac{4}{\sqrt{|K|}}\bigg)w_1+(h_2-2)w_2\bigg\}}dx.\nonumber
\end{align}

To prove the existence of a solution of the system \eqref{w1}-\eqref{w2}, it is sufficient to show that the functional $I$, defined by \eqref{functional2}, attains a unique interior critical point in some open ball in $H^1_2$. It is straightforward to show that the functional $I$ is strictly convex, Gateaux differentiable, and hence weakly lower semi-continuous. To prove the existence of an interior critical point, we just need to show that the Gateaux derivative of $I$ satisfies a coercive lower bound. We begin by rewriting the functional $I$ in the form
\begin{align}
I(&w_1,w_2)=\frac{1}{2}||\nabla w_1||^2_2+\dfrac{1}{2}||\nabla w_2||^2_2+\frac{2k_{11}}{|K|}\bigg(e^{u_0'},e^{\sqrt{|K|}w_1}-1-\sqrt{|K|}w_1\bigg)_2\\
&+\frac{2k_{11}}{|K|}\bigg(e^{u_0''},e^{(k_{21}\sqrt{|K|}w_1+|K|w_2)/k_{11}}-1-\frac{1}{k_{11}}(k_{21}\sqrt{|K|}w_1+|K|w_2)\bigg)_2\nonumber\\
&+\bigg(w_1,h_1+\frac{2k_{11}}{\sqrt{|K|}}(e^{u_0'}-1)+\frac{2k_{12}}{\sqrt{|K|}}(e^{u_0''}-1)\bigg)_2\nonumber\\
&+\bigg(w_2,h_2+2(e^{u_0''}-1)\bigg)_2,\nonumber
\end{align}
where  $(\cdot,\cdot)_2$ denotes the inner product over $L^2(\mathbb{R}^2)$.
Calculating the Gateaux derivative of $I$, denoted by $dI=dI(w;w)$, we get
\begin{align}
&dI-\sum_{i=1}^{2}||\nabla w_i||^2_2= \dfrac{2k_{11}}{|K|}\bigg(\sqrt{|K|}w_1,e^{u_0'+\sqrt{|K|}w_1}-1+\dfrac{\sqrt{|K|}}{2k_{11}}h_1-\frac{k_{21}}{2k_{11}}h_2\bigg)_2\\
&+\dfrac{2k_{11}}{|K|}\bigg(\frac{1}{k_{11}}(k_{21}\sqrt{|K|}w_1+|K|w_2),e^{u_0''+(k_{21}\sqrt{|K|}w_1+|K|w_2)/k_{11}}-1+ \dfrac{1}{2}h_2\bigg)_2.\nonumber
\end{align}

\begin{lem} For the undetermined parameter $\mu>0$ sufficiently large, there exists positive constants $C_1$, $C_2$, and $C_3$ such that 
\begin{align}
dI(w;w)-\sum_{i=1}^{2}||\nabla w_i||^2_2&\geq C_1\int_{\R^2}{\dfrac{(\sqrt{|K|}w_1)^2}{1+|\sqrt{|K|}w_1|}}dx\label{eq12}\\
&+C_2\int_{\R^2}{\dfrac{\bigg(\dfrac{1}{k_{11}}(k_{21}\sqrt{|K|}w_1+|K|w_2)\bigg)^2}{1+\dfrac{1}{k_{11}}|k_{21}\sqrt{|K|}w_1+|K|w_2|}}dx-C_3\nonumber.
\end{align}
\end{lem}
\textbf{Proof}. We note that the inner product terms of $dI$ are of the general form,
$$\alpha\bigg(v,e^{u_0+v}-1+g\bigg)_2,$$ for some positive constant $\alpha$. Define 
\begin{align*}
M(v)=\alpha\bigg(v,e^{u_0+v}-1+g\bigg)_2
\end{align*}
and let $v=v^+-v^-$ where $v^+=\max\{0,v\}$ and $v^-=\max\{0,-v\}$. So $M(v)=M(v^+)+M(-v^-)$.

From the elementary inequalities, $e^x\geq x+1$ and $xy\geq-\dfrac{1}{2}(x^2+y^2)$ for all $x,y\in\R$, we get 
\begin{align*}
M(v^+)&\geq\dfrac{\alpha}{2}\int_{\R^2}{(v^+)^2}dx-\dfrac{\alpha}{2}\int_{\R^2}{(u_0+g)^2}dx
\geq\dfrac{\alpha}{2}\int_{\R^2}{(v^+)^2}dx-\beta,
\end{align*}
where $\beta$ is a positive constant, since $u_0$ and $g$ belong in $L^2(\R^2)$. For any $x\geq 0$ we note that $x^2\geq \dfrac{x^2}{1+x}$, therefore,
\begin{equation}
M(v^+)\geq\dfrac{\alpha}{2}\int_{\R^2}{\dfrac{(v^+)^2}{1+v^+}}dx-\beta.\label{eq9}
\end{equation}
Let us now consider the equation 
\begin{align*}
&M(-v^-)=\alpha\bigg(-v^-,e^{u_0-v^-}-1+g\bigg)_2
=\alpha\bigg(v^-,1-g-e^{u_0-v^-}\bigg)_2.
\end{align*}
From the elementary inequality $1-e^{-x}\geq\dfrac{x}{1+x}$ for any $x\geq 0$, it follows that
\begin{equation*}
\alpha v^-(1-g-e^{u_0-v^-})
\geq\alpha\dfrac{(v^-)^2}{1+v^-}(1-g)+\alpha\dfrac{v^-}{1+v^-}(1-g-e^{u_0}).
\end{equation*}
For $\mu>0$ large enough, we may obtain $1-g>\dfrac{1}{2}$. In addition, since $1-e^{u_0}$ and $g$ are in $L^2(\R^2)$, we have
\begin{equation*}
\bigg|\int_{\R^2}{\dfrac{v^-}{1+v^-}(1-g-e^{u_0})}dx\bigg|\leq\dfrac{1}{4}\int_{\R^2}{\dfrac{(v^-)^2}{1+v^-}}dx+\int_{\R^2}{(1-g-e^{u_0})^2}dx.
\end{equation*}
From the absolute value we can conclude that
\begin{equation*}
\int_{\R^2}{\dfrac{v^-}{1+v^-}(1-g-e^{u_0})}dx
\geq-\dfrac{1}{4}\int_{\R^2}{\dfrac{(v^-)^2}{1+v^-}}dx-\tilde{\beta},
\end{equation*}
for some positive constant $\tilde{\beta}$, since $1-g-e^{u_0}$ belongs in $L^2(\R^2)$. Thus, we have
\begin{equation}
M(-v^-)\geq\dfrac{\alpha}{4}\int_{\R^2}{\dfrac{(v^-)^2}{1+v^-}}dx-\tilde{\beta}.\label{eq10}
\end{equation}
Consequently, \eqref{eq9} and \eqref{eq10} gives us
\begin{equation}
M(v)=M(v^+)+M(-v^-)\geq C_1\int_{\R^2}{\dfrac{v^2}{1+|v|}}dx-C_2,\label{eq15}
\end{equation}
for some positive constants $C_1$ and $C_2$. Applying \eqref{eq15} to each inner product term of $dI$ gives us the desired inequality \eqref{eq12}.$\quad\square$

Absorbing constants into $C_1$ and $C_2$, we may rewrite \eqref{eq12} in the cleaner form
\begin{align}
dI(w;w)-\sum_{i=1}^{2}||\nabla w_i||^2_2&\geq C_1\int_{\R^2}{\dfrac{w_1^2}{(1+|w_1|)^2}}dx\label{eq16}\\
&+C_2\int_{\R^2}{\dfrac{(k_{21}\sqrt{|K|}w_1+|K|w_2)^2}{(1+|k_{21}\sqrt{|K|}w_1+|K|w_2|)^2}}dx-C_3.\nonumber
\end{align}
We reuse the positive constants $C_0,\ldots, C_3$ and keep in mind that the importance of these constants is that they remain positive, of finite value, and independent of $w_1$ and $w_2$.

Define the transformation $T(w_1,w_2)=(u_1,u_2)$ such that 
\begin{equation}
u_1=w_1\qquad \text{and}\qquad
u_2=k_{21}\sqrt{|K|}w_1+|K|w_2.
\end{equation}
Applying this transformation to \eqref{eq16}, we get
\begin{equation*}
dI(w;w)-\sum_{i=1}^{2}||\nabla w_i||^2_2\geq C_1\int_{\R^2}{\dfrac{u_1^2}{(1+|u_1|)^2}}dx+C_2\int_{\R^2}{\dfrac{u_2^2}{(1+|u_2|)^2}}dx-C_3.
\end{equation*}
Since $T$ is an invertible transformation, there is a positive constant $C_0$ such that 
\begin{equation*}
\sum\limits_{i=1}^{2}{||\nabla w_i||_2^2}\geq C_0 \sum\limits_{i=1}^{2}{||\nabla u_i||_2^2}.
\end{equation*}
Thus, we have the lower bound
\begin{equation*}
dI(w;w)\geq C_0\sum_{i=1}^{2}||\nabla u_i||^2_2+C_1\int_{\R^2}{\dfrac{u_1^2}{(1+|u_1|)^2}}dx\\
+C_2\int_{\R^2}{\dfrac{u_2^2}{(1+|u_2|)^2}}dx-C_3,
\end{equation*}
which can be rewritten as
\begin{equation*}
dI(w;w)\geq C_0 \sum_{i=1}^{2}||\nabla u_i||^2_2
+ C_1\sum_{i=1}^{2}\bigg(1+||\nabla u_i||^2_2+\int_{\R^2}{\dfrac{u_i^2}{(1+|u_i|)^2}}dx\bigg)
-C_2.
\end{equation*}
Using the following standard interpolation inequality over $W^{1,2}(\R^2)$:
\begin{equation*}
\int_{\mathbb{R}^2}{|v|^4}dx\leq  2\int_{\mathbb{R}^2}{|v|^2}dx\int_{\mathbb{R}^2}{|\nabla v|^2}dx,\qquad v\in W^{1,2}(\mathbb{R}^2),
\end{equation*}
we may write
\begin{equation*}
\bigg(\int_{\mathbb{R}^2}{|v|^2}dx\bigg)^2\leq 4\int_{\mathbb{R}^2}{\dfrac{|v|^2}{(1+|v|)^2}}dx\int_{\mathbb{R}^2}{|v|^2}dx\bigg(1+\int_{\mathbb{R}^2}{|\nabla v|^2}dx\bigg).
\end{equation*}
Recall the generalized arithmetic and geometric inequality, $\prod\limits_{i=1}^{n}{a_i^{\lambda_i}}\leq \sum\limits_{i=1}^{n}{\lambda_i a_i}$,
where the $a_i$'s and $\lambda_i$'s are non-negative real numbers such that $\sum\limits_{i=1}^{n}{\lambda_i}=1$. As a special case, we get 
\begin{align*}
\bigg(\int_{\mathbb{R}^2}{|v|^2}dx\bigg)^2
\leq&C\bigg(1+\bigg[\int_{\mathbb{R}^2}{\dfrac{|v|^2}{(1+|v|)^2}}dx\bigg]^4+\bigg[\int_{\mathbb{R}^2}{|\nabla v|^2}dx\bigg]^4\bigg)\\
&+\dfrac{1}{2}\bigg(\int_{\mathbb{R}^2}{|v|^2}dx\bigg)^2.\nonumber
\end{align*} 
It follows that, 
\begin{equation}
\bigg(\int_{\mathbb{R}^2}{|v|^2}dx\bigg)^{\frac{1}{2}}\leq C\bigg(1+\int_{\mathbb{R}^2}{\dfrac{|v|^2}{(1+|v|)^2}}dx+\int_{\mathbb{R}^2}{|\nabla v|^2}dx\bigg)\label{eq18}
\end{equation}
for some positive constant $C$. By \eqref{eq18}, we obtain 
\begin{equation}
dI(w;w)\geq C_0 \sum_{i=1}^{2}||\nabla u_i||^2_{L^2(\mathbb{R}^2)}+||u_i||_{L^2(\mathbb{R}^2)} -C_1.
\end{equation}
Using the relationship between $u_i$ and $w_i$, we get the coercive lower bound 
\begin{equation}
dI(w;w)\geq C_0 \bigg(||w_1||_{1,2}+||w_2||_{1,2}\bigg) -C_1,\label{eq20}
\end{equation}
where $C_0$ and $C_1$ are positive constants independent of $w_1,w_2\in W^{1,2}(\mathbb{R}^2)$. From \eqref{eq20}, we conclude that, for any $\delta>0$ there is an $0<R<\infty$ so that 
\begin{equation}
\inf\limits_{||w||_{1,2}=R}{dI(w;w)}\geq\delta,
\end{equation}
which gives the existence of an interior critical point of $I$ in some open ball in $H^1_2$. Therefore, the existence part of Theorem \ref{fullPlane} is established. The uniqueness follows from the strict convexity of $I$.

\section{Asymptotic Analysis and Quantized Magnetic Flux}
In this section, we prove the exponential decay estimates of Theorem \ref{fullPlane}. We also establish the quantized magnetic flux integral formulas of Theorem \ref{quantizedInt}, over the full plane. Our analysis follows similar ideas as in \cite{Taubes, Yang1, Yang4}.
\begin{lem}\label{lem1}
Let $w=(w_1,w_2)$ in $H^1_2$ be the solution of the variational equations \eqref{w1}-\eqref{w2}. Then $w_i(x)\rightarrow 0$ as $|x|\rightarrow \infty$, for $i=1,2$. More explicitly, $w_i$ satisfies the uniform decay 
\begin{equation}
\lim_{R\rightarrow\infty}{\sup_{|x|=R}|w_i(x)|}=0.
\end{equation}
\end{lem}
\textbf{Proof}. We will need the following embedding inequality \cite{Taubes} for $p>2$ , 
\begin{equation}
||f||_p\leq\bigg(\pi\bigg[\dfrac{p}{2}-1\bigg]\bigg)^{\frac{p-2}{2p}}||f||_{1,2}.\label{5.1}
\end{equation}
Let us first justify that $e^f-1\in L^2(\mathbb{R}^2)$ for any $f$ in $W^{1,2}(\mathbb{R}^2)$. By expanding $(e^f-1)^2$ and using the Taylor series for $e^f$ we obtain 
\begin{equation}
(e^f-1)^2=f^2+\sum\limits_{k=3}^{\infty}\dfrac{2^k-2}{k!}f^k.\label{5.2}
\end{equation}
Integrating \eqref{5.2} over $\mathbb{R}^2$ and using the embedding inequality \eqref{5.1} we get
\begin{align}
||e^f-1||_2^2&=||f||^2_2+\sum\limits_{k=3}^{\infty}\dfrac{2^k-2}{k!}\int{f^k}dx\label{5.3}\\
&\leq||f||^2_2+\sum\limits_{k=3}^{\infty}\dfrac{2^k-2}{k!}\bigg(\pi\bigg[\dfrac{k}{2}-1\bigg]\bigg)^{\frac{k-2}{2}}||f||^k_{1,2},\nonumber
\end{align}
which is a convergent series. Therefore, $e^f-1$ is in $L^2(\mathbb{R}^2)$ for any $f$ in $W^{1,2}(\mathbb{R}^2)$.

Let $w_i$, in $W^{1,2}(\mathbb{R}^2)$, be the solutions to \eqref{w1}-\eqref{w2}, then they are also in $L^p(\mathbb{R}^2)$ for $p>2$ by \eqref{5.1}. In particular, the $w_i$ are in $L^{\infty}(\R^2)$. To establish the desire uniform decay, it is sufficient to show that $w_i$ belongs to $W^{2,2}(\mathbb{R}^2)$. To this end, we rewrite the system \eqref{w1}-\eqref{w2} in the form 
\begin{align}
\Delta w_1&=\dfrac{2K_{11}}{\sqrt{|K|}}(e^{u_0'}-1)e^{\sqrt{|K|}w_1}+\dfrac{2K_{11}}{\sqrt{|K|}}(e^{\sqrt{|K|}w_1}-1)\label{5.4}\\
&+\dfrac{2K_{12}}{\sqrt{|K|}}(e^{u_0''}-1)e^{\frac{1}{K_{11}}(|K|w_2+K_{21}\sqrt{|K|}w_1)}\nonumber\\
&+\dfrac{2K_{12}}{\sqrt{|K|}}(e^{\frac{1}{K_{11}}(|K|w_2+K_{21}\sqrt{|K|}w_1)}-1)+h_1\nonumber\\
\Delta w_2&=2(e^{u_0''}-1)e^{\frac{1}{K_{11}}(|K|w_2+K_{21}\sqrt{|K|}w_1)}\label{5.5}\\
&+2(e^{\frac{1}{K_{11}}(|K|w_2+K_{21}\sqrt{|K|}w_1)}-1)+h_2,\nonumber
\end{align}
and show that the right hand sides of \eqref{5.4}-\eqref{5.5} are in $L^2(\mathbb{R}^2)$.

We note that $h_1$ and $h_2$ are in $L^2(\mathbb{R}^2)$. The terms, $e^{\sqrt{|K|}w_1}-1$ and $e^{(|K|w_2+k_{21}\sqrt{|K|}w_1)/\kappa_{11}}-1$ are in $L^2{(\mathbb{R}^2)}$ by \eqref{5.3}. Moreover, the term $(e^{u_0'}-1)e^{\sqrt{|K|}w_1}$ is in $L^2{(\mathbb{R}^2)}$, by
\begin{align}
\int{\bigg((e^{u_0'}-1)e^{\sqrt{|K|}w_1}\bigg)^2}dx&=\int{(e^{u_0'}-1)^2e^{2\sqrt{|K|}w_1}}dx\\
&\leq M \int{(e^{u_0'}-1)^2}dx<\infty,\nonumber
\end{align}
where $M=\sup\limits_{x\in\mathbb{R}^2}e^{2\sqrt{|K|}w_1(x)}<\infty$ from $w_1\in L^{\infty}(\mathbb{R}^2)$ and $e^{u_0'}-1\in L^2(\mathbb{R}^2)$. Similarly, we can conclude that $(e^{u_0''}-1)e^{(|K|w_2+k_{21}\sqrt{|K|}w_1)/k_{11}}$ is in $L^2(\mathbb{R}^2)$. Thus, the uniform decay of $w_1$ and $w_2$ is established. $\quad\square$

\begin{lem} Let $w=(w_1,w_2)$ be as stated in Lemma \ref{lem1}. Then $\nabla w_1$ and $\nabla w_2$, also satisfy the uniform decay $|\nabla w_1|,|\nabla w_2|\rightarrow 0$ as $|x|\rightarrow \infty$.
\end{lem}
\textbf{Proof}. We note that $w_i$ belongs in $W^{1,p}(\mathbb{R}^2)$ for all $p\geq 2$ and $i=1,2$. We simply need to extend $w_i$ to belong to $W^{2,p}(\mathbb{R}^2)$ for all $p> 2$. To achieve this, it is sufficient to show that the $\Delta w_i$ is in $L^p(\mathbb{R}^2)$ for $p>2$. This follows by proving that each term on the right hand side of \eqref{5.4}-\eqref{5.5} is in $L^p(\mathbb{R}^2)$ for $p>2$. It is enough to show that $\nabla(e^{(|K|w_2+k_{21}\sqrt{|K|}w_1)/k_{11}}-1)$ and $\nabla (e^{\sqrt{|K|}w_1}-1)$ are in $L^2(\mathbb{R}^2)$.

In general, if $f$ is in $W^{1,p}(\mathbb{R}^2)$ for all $p\geq 2$, we can conclude that
\begin{align*}
\int{|\nabla(e^f-1)|^2}dx&=\int{e^{2f}|\nabla f|^2}dx\\&\leq M\int{|\nabla f|^2}dx<\infty,
\end{align*}
where $M=\sup\limits_{x\in\mathbb{R}^2}e^{2f(x)}<\infty$ since $f$ is in $L^{\infty}(\mathbb{R}^2)$ and $\nabla f$ is in $L^2(\mathbb{R}^2)$. By \eqref{5.1},  $e^{(|K|w_2+k_{21}\sqrt{|K|}w_1)/k_{11}}-1$ and $e^{\sqrt{|K|}w_1}-1$ are in $L^p(\mathbb{R}^2)$ for all $p>2$. The term $(e^{u_0'}-1)e^{\sqrt{|K|}w_1}$ is in $L^p{(\mathbb{R}^2)}$ for $2\leq p<\infty$ since 
\begin{align*}
\int{\bigg((e^{u_0'}-1)e^{\sqrt{|K|}w_1}\bigg)^p}dx&=\int{(e^{u_0'}-1)^pe^{p\sqrt{|K|}w_1}}dx\\&\leq M \int{(e^{u_0'}-1)^2}dx<\infty,
\end{align*}
where $M=\sup\limits_{x\in\mathbb{R}^2}e^{p\sqrt{|K|}w_1(x)}<\infty$ since $w_1\in L^{\infty}(\mathbb{R}^2)$ and $e^{u_0'}-1\in L^p(\mathbb{R}^2)$ for $p\geq 1$. Moreover, $(e^{u_0'}-1)e^{\sqrt{|K|}w_1}$ is in $L^{\infty}{(\mathbb{R}^2)}$ since $|e^{u_0'}-1|\leq 1$ and $w_1$ is in  $L^{\infty}{(\mathbb{R}^2)}$. Hence, $(e^{u_0'}-1)e^{\sqrt{|K|}w_1}$ is in $L^p{(\mathbb{R}^2)}$ for $p\geq 2$. 

Similarly, we can show that $(e^{u_0''}-1)e^{(|K|w_2+k_{21}\sqrt{|K|}w_1)/k_{11}}$ is in $L^p(\mathbb{R}^2)$ for $p\geq 2$. Therefore, the uniform decay of $\nabla w_1$ and $\nabla w_2$ follows. $\quad\square$

From $(w_1,w_2)$, defined in Lemma \ref{lem1}, we get $(\tilde{v}_1,\tilde{v}_2)$ by the transformation \eqref{Choleski}, and hence a solution pair $(u_1,u_2)$ is obtained as a solution of \eqref{elliptic1}-\eqref{elliptic2} on the full plane $\R^2$, satisfying \eqref{topobound}. To complete Theorem \ref{fullPlane}, we just need to establish the exponential decay estimates \eqref{2.46} and \eqref{2.47}.

\begin{lem}\label{lem3}
For the pair $(u_1,u_2)$ stated above, there holds the exponential decay estimate
\begin{equation}
(u_1+\ln 2)^2+(u_2+\ln 2)^2\leq C_{\epsilon}e^{-(1-\epsilon)\sqrt{\lambda_0}|x|},\label{5.7}
\end{equation} 
for $|x|$ sufficiently large, $\epsilon\in(0,1)$ arbitrary, $C_{\epsilon}>0$ a constant depending on $\epsilon$, and 
$\lambda_0= 4 \min \left\{ 2,\frac{2q}{p}\right\}$.
\end{lem}
\textbf{Proof}. The eigenvalues of the matrix $K$ are $\lambda_1=2$ and $\lambda_2=\frac{2q}{p}$. Both eigenvalues are greater than zero and since $p\neq q$, we have that $\lambda_1\neq\lambda_2$. Hence, there is an orthogonal $2\times 2$ matrix $\mathcal{O}$ that diagonalizes $K$, i.e.,
\begin{displaymath}
\mathcal{O}^{-1} K \mathcal{O}=
\left(\begin{array}{cc}
\lambda_1 & 0\\
0 & \lambda_2
\end{array}\right):=\Lambda.
\end{displaymath}
Take $R_0>\max\{|p_1|,\dots,|p_{N_1}|,|q_1|,\dots,|q_{N_2}|\}$. Outside the disk of radius $R_0$ centered at the origin, $D_{R_0}=\big\{x\in\mathbb{R}^2\big| |x|\leq R_0\big\}$, the system \eqref{elliptic1}-\eqref{elliptic2} considered over the full plane, becomes 
\begin{equation}
\Delta u_i=4\left(k_{i1}e^{u_1}+k_{i2}e^{u_2}-1\right)\qquad\text{where}\quad i=1,2.\label{5.8}
\end{equation}
Rewrite \eqref{5.8} in the form, 
\begin{equation}
\Delta u_i=4k_{i1}u_1+4k_{i2}u_2+2k_{i1}(2e^{u_1}-2u_1-1)+2k_{i2}(2e^{u_2}-2u_2-1).\label{5.9}
\end{equation}
We would like the terms $2e^{u_i}-2u_i-1$ to converge to 0 as $|x|\rightarrow \infty$. Thus we define $v_i=u_i+\ln(2)$. It then follows that each $v_i\rightarrow 0$ as $|x|\rightarrow\infty$ in the sense of Lemma \ref{lem1}. Then \eqref{5.9} becomes
\begin{equation}
\Delta v_i=2k_{i1}v_1+2k_{i2}v_2+2K_{i1}(e^{v_1}-v_1-1)+2k_{i2}(e^{v_2}-v_2-1).\label{5.10}
\end{equation}
Define the new variables $U_1$ and $U_2$ such that
\begin{equation*}
\left(\begin{array}{c}
U_1\\
U_2
\end{array}\right)=\mathcal{O}
\left(\begin{array}{c}
v_1\\
v_2
\end{array}\right).
\end{equation*}
Using the variables $U_1$ and $U_2$, we express \eqref{5.10} in the form
\begin{equation*}
\left(\begin{array}{c}
\Delta U_1\\
\Delta U_2
\end{array}\right)=
\left(\begin{array}{c}
2\lambda_1 U_1\\
2\lambda_2 U_2
\end{array}\right)
+
2\mathcal{O}
\left(\begin{array}{c}
e^{\frac{1}{\sqrt{2}}(U_1+U_2)}-\frac{1}{\sqrt{2}}(U_1+U_2)-1\\
e^{\frac{1}{\sqrt{2}}(U_1-U_2)}-\frac{1}{\sqrt{2}}(U_1-U_2)-1
\end{array}\right).
\end{equation*}
By using the Taylor expansion of $e^x$, we may write
\begin{equation*}
\Delta U_i=2\lambda_i U_i + a_{i1}(U_1,U_2)U_1+a_{i2}(U_1,U_2)U_2,
\end{equation*}
where $a_{jk}(U_1,U_2)\rightarrow 0$ as $|x|\rightarrow \infty$ $(j,k=1,2)$. In two dimensions, we recall the inequality 
\begin{equation*}
\Delta f^2=2f\Delta f +2\left(\dfrac{\partial f}{\partial x}\right)^2+2\left(\dfrac{\partial f}{\partial y}\right)^2\geq 2f\Delta f.
\end{equation*}
The following inequality then follows, 
\begin{align*}
\Delta (U_1^2+U_2^2)&\geq \lambda_0(U_1^2+ U_2^2)-a(U_1,U_2)(U_1^2+ U_2^2),
\end{align*}
where the function $a(U_1,U_2)\rightarrow 0$ as $|x|\rightarrow\infty$. Consequently, for any $\epsilon\in(0,1)$ we can find an $R>R_0$ large enough, so that 
\begin{equation}
\Delta (U_1^2+U_2^2)\geq \left(1-\frac{\epsilon}{2}\right)\lambda_0(U_1^2+ U_2^2),\qquad |x|>R.\label{5.11}
\end{equation}
Introduce the following comparison function, 
\begin{equation}
\xi(x)=Ce^{-\sigma|x|},\qquad |x|>0,\qquad C,\sigma\in\mathbb{R},\qquad C,\sigma>0.\label{5.12}
\end{equation}
Then,
\begin{equation}
\Delta\xi=\sigma^2\xi-\dfrac{\sigma\xi}{|x|}.\label{5.13}
\end{equation}
Subtracting \eqref{5.11} and \eqref{5.13}, for $|x|>R$, we get
\begin{align*}
\Delta (U_1^2+U_2^2-\xi)&=\Delta (U_1^2+U_2^2)-\Delta\xi\\
&\geq \left(1-\frac{\epsilon}{2}\right)\lambda_0(U_1^2+ U_2^2)-\left(\sigma^2\xi-\dfrac{\sigma\xi}{|x|}\right)\\
&\geq \left(1-\frac{\epsilon}{2}\right)\lambda_0(U_1^2+ U_2^2)-\sigma^2\xi.
\end{align*}
We have the freedom to select $\sigma^2=\left(1-\frac{\epsilon}{2}\right)\lambda_0$, which gives us 
\begin{equation}
\Delta (U_1^2+U_2^2-\xi)\geq \sigma^2(U_1^2+ U_2^2-\xi)\quad\text{for }\quad |x|>R.
\end{equation}
Now we can select $C$ in \eqref{5.12} large enough so that $U_1^2+ U_2^2-\xi\leq 0$ for $|x|=R$. Let's denote $C$ by $C_{\epsilon}$ to point out its dependence on $\epsilon$. We would like to extend this inequality so that it holds for all $|x|\geq R$. Since $U_1^2+ U_2^2\rightarrow 0$ as $|x|\rightarrow\infty$ and applying the maximum principle, we can conclude that  $U_1^2+ U_2^2-\xi\leq 0$ for $|x|\geq R$.
For any $\epsilon\in(0,1)$ we have the following useful inequality $\sqrt{1-\epsilon/2}>1-\epsilon$.
Hence, for $|x|\geq R$ we get
\begin{equation*}
U_1^2+U_2^2\leq C_{\epsilon}e^{-\sqrt{(1-\epsilon/2)\lambda_0}|x|}
\leq C_{\epsilon}e^{-(1-\epsilon)\sqrt{\lambda_0}|x|}.
\end{equation*}
By the orthogonality of $\mathcal{O}$, the equation  $U_1^2+U_2^2=v_1^2+v_2^2$ follows. Therefore, for $|x|\geq R$, we have the desire inequality
\begin{equation*}
v_1^2+v_2^2=(u_1+\ln 2)^2+(u_2+\ln 2)^2
\leq C_{\epsilon}e^{-(1-\epsilon)\sqrt{\lambda_0}|x|}.\qquad\square
\end{equation*}

\begin{lem}
$u_1$ and $u_2$, from Lemma \ref{lem3}, also satisfy the inequality
\begin{equation*}
|\nabla u_1|^2+|\nabla u_2|^2\leq C_{\epsilon}e^{-(1-\epsilon)\sqrt{\lambda_0}|x|},
\end{equation*}
where $\epsilon$, $C_{\epsilon}$, and $\lambda_0$ are as defined in Lemma \ref{lem3}.
\end{lem}
\textbf{Proof}. Take $R$ as defined in Lemma \ref{lem3}. Differentiating equations \eqref{5.10}, for $|x|>R$, we get
\begin{equation}
\Delta \partial_j v_i=2k_{i1}\partial_jv_1+2k_{i2}\partial_jv_2+2k_{i1}(e^{v_1}-1)\partial_jv_1+2k_{i2}(e^{v_2}-1)\partial_jv_2,\label{5.16}
\end{equation}
for $i,j=1,2$ and $\partial_j\equiv\frac{\partial}{\partial x_j}$. 

Define  $U=(U_1,U_2)^{\tau}=(\partial_1 v_1 \partial_2 v_2)^{\tau}$ and $E(x)=\mbox{diag}\{e^{u_1(x)},e^{u_2(x)}\}$, where $\mbox{diag}\{a,b\}$ is a $2\times 2$ diagonal matrix with diagonal entries $a$ and $b$, respectively. Then the system \eqref{5.16} may be rewritten in the form
\begin{equation*}
\Delta U=2KU+2K(E(x)-I_2)U,
\end{equation*}
where $I_2$ is the $2\times 2$ identity matrix. Consequently, we can establish the following inequality for $|x|>R$,
\begin{align*}
\Delta |U|^2 &\geq\lambda_0 |U|^2-b(U_1,U_2)|U|^2,
\end{align*}
where $b(U_1,U_2)\rightarrow 0$ as $|x|\rightarrow\infty$. Hence, as in Lemma \ref{lem3}, for $|x|\geq R$, we arrived at the inequality
\begin{equation*}
U_1^2+U_2^2=(\partial_j v_1)^2+(\partial_j v_2)^2 \leq C_{\epsilon,j}e^{-(1-\epsilon)\sqrt{\lambda_0}|x|},
\end{equation*}
where $C_{\epsilon,j}$ are positive constant depending on $\epsilon$. Therefore, we can take $C_{\epsilon}=2\max\{C_{\epsilon,1},C_{\epsilon,2}\}$ and obtain the desire inequality
\begin{equation*}
|\nabla u_1|^2+|\nabla u_2|^2\leq C_{\epsilon}e^{-(1-\epsilon)\sqrt{\lambda_0}|x|}\qquad\text{for}\qquad |x|\geq R. \qquad\square
\end{equation*}

As a result of the exponential decay estimates, we get the quantized magnetic flux integrals. A direct calculation shows that the integrals of the functions $g_0'(x)$ and $g_0''(x)$ over $\mathbb{R}^2$ are independent of the parameter $\mu$. More explicitly,
\begin{equation*}
\int_{\mathbb{R}^2}{g_0'(x)}dx=4\pi N_1\qquad \text{and}\qquad \int_{\mathbb{R}^2}{g_0''(x)}dx=4\pi N_2.
\end{equation*}
The divergence theorem in two dimensions gives  
\begin{equation*}
\int_{\mathbb{R}^2}{\Delta \tilde{v}_1}dx=\int_{\mathbb{R}^2}{\Delta \tilde{v}_2}dx=0.
\end{equation*}
By integrating equations \eqref{eq7}-\eqref{eq8} over the full plane, we get
\begin{align*}
\int_{\mathbb{R}^2}{(k_{i1}e^{u_1}+k_{i2}e^{u_2}-1)}dx&=-\pi N_i.
\end{align*}

Similarly to the doubly periodic case, we obtain the quantized magnetic flux integral formulas of Theorem \ref{quantizedInt}, 
\begin{equation*}
\int_{\R^2}{B_{12}}dx=-2\pi p N_1\qquad\text{and}\qquad
\int_{\R^2}{\tilde{B}_{12}}dx=-2\pi pN_2.
\end{equation*} 

% ------------------------------------------------------------------------
%\nocite{*}
\def\bibsection{\section*{\refname}} 
\bibliography{Medina_FQHE_Rev_9_10_15}% Produces the bibliography via BibTeX.

%merlin.mbs apsrev4-1.bst 2010-07-25 4.21a (PWD, AO, DPC) hacked
%Control: key (0)
%Control: author (8) initials jnrlst
%Control: editor formatted (1) identically to author
%Control: production of article title (-1) disabled
%Control: page (0) single
%Control: year (1) truncated
%Control: production of eprint (0) enabled
\providecommand{\noopsort}[1]{}\providecommand{\singleletter}[1]{#1}%
\begin{thebibliography}{38}%
\makeatletter
\providecommand \@ifxundefined [1]{%
 \@ifx{#1\undefined}
}%
\providecommand \@ifnum [1]{%
 \ifnum #1\expandafter \@firstoftwo
 \else \expandafter \@secondoftwo
 \fi
}%
\providecommand \@ifx [1]{%
 \ifx #1\expandafter \@firstoftwo
 \else \expandafter \@secondoftwo
 \fi
}%
\providecommand \natexlab [1]{#1}%
\providecommand \enquote  [1]{``#1''}%
\providecommand \bibnamefont  [1]{#1}%
\providecommand \bibfnamefont [1]{#1}%
\providecommand \citenamefont [1]{#1}%
\providecommand \href@noop [0]{\@secondoftwo}%
\providecommand \href [0]{\begingroup \@sanitize@url \@href}%
\providecommand \@href[1]{\@@startlink{#1}\@@href}%
\providecommand \@@href[1]{\endgroup#1\@@endlink}%
\providecommand \@sanitize@url [0]{\catcode `\\12\catcode `\$12\catcode
  `\&12\catcode `\#12\catcode `\^12\catcode `\_12\catcode `\%12\relax}%
\providecommand \@@startlink[1]{}%
\providecommand \@@endlink[0]{}%
\providecommand \url  [0]{\begingroup\@sanitize@url \@url }%
\providecommand \@url [1]{\endgroup\@href {#1}{\urlprefix }}%
\providecommand \urlprefix  [0]{URL }%
\providecommand \Eprint [0]{\href }%
\providecommand \doibase [0]{http://dx.doi.org/}%
\providecommand \selectlanguage [0]{\@gobble}%
\providecommand \bibinfo  [0]{\@secondoftwo}%
\providecommand \bibfield  [0]{\@secondoftwo}%
\providecommand \translation [1]{[#1]}%
\providecommand \BibitemOpen [0]{}%
\providecommand \bibitemStop [0]{}%
\providecommand \bibitemNoStop [0]{.\EOS\space}%
\providecommand \EOS [0]{\spacefactor3000\relax}%
\providecommand \BibitemShut  [1]{\csname bibitem#1\endcsname}%
\let\auto@bib@innerbib\@empty
%</preamble>
\bibitem [{\citenamefont {Han}\ and\ \citenamefont {Song}(2011)}]{JHan}%
  \BibitemOpen
  \bibfield  {author} {\bibinfo {author} {\bibfnamefont {J.}~\bibnamefont
  {Han}}\ and\ \bibinfo {author} {\bibfnamefont {K.}~\bibnamefont {Song}},\
  }\href@noop {} {\bibfield  {journal} {\bibinfo  {journal} {Nonlinear
  Analysis}\ }\textbf {\bibinfo {volume} {74}},\ \bibinfo {pages} {7426}
  (\bibinfo {year} {2011})}\BibitemShut {NoStop}%
\bibitem [{\citenamefont {Han}\ and\ \citenamefont {Yang}(2015)}]{Han}%
  \BibitemOpen
  \bibfield  {author} {\bibinfo {author} {\bibfnamefont {X.}~\bibnamefont
  {Han}}\ and\ \bibinfo {author} {\bibfnamefont {Y.}~\bibnamefont {Yang}},\
  }\href@noop {} {\bibfield  {journal} {\bibinfo  {journal} {Commun. Math.
  Phys.}\ }\textbf {\bibinfo {volume} {333}},\ \bibinfo {pages} {229} (\bibinfo
  {year} {2015})}\BibitemShut {NoStop}%
\bibitem [{\citenamefont {Lieb}\ and\ \citenamefont {Yang}(2012)}]{Lieb}%
  \BibitemOpen
  \bibfield  {author} {\bibinfo {author} {\bibfnamefont {E.}~\bibnamefont
  {Lieb}}\ and\ \bibinfo {author} {\bibfnamefont {Y.}~\bibnamefont {Yang}},\
  }\href@noop {} {\bibfield  {journal} {\bibinfo  {journal} {Commun. Math.
  Phys.}\ }\textbf {\bibinfo {volume} {313}},\ \bibinfo {pages} {445} (\bibinfo
  {year} {2012})}\BibitemShut {NoStop}%
\bibitem [{\citenamefont {Lin}\ \emph {et~al.}(2013)\citenamefont {Lin},
  \citenamefont {Tarantello},\ and\ \citenamefont {Yang}}]{CSLin}%
  \BibitemOpen
  \bibfield  {author} {\bibinfo {author} {\bibfnamefont {C.}~\bibnamefont
  {Lin}}, \bibinfo {author} {\bibfnamefont {G.}~\bibnamefont {Tarantello}}, \
  and\ \bibinfo {author} {\bibfnamefont {Y.}~\bibnamefont {Yang}},\ }\href@noop
  {} {\bibfield  {journal} {\bibinfo  {journal} {J. Diff. Eqs.}\ }\textbf
  {\bibinfo {volume} {254}},\ \bibinfo {pages} {1437} (\bibinfo {year}
  {2013})}\BibitemShut {NoStop}%
\bibitem [{\citenamefont {Nam}(2013)}]{Nam}%
  \BibitemOpen
  \bibfield  {author} {\bibinfo {author} {\bibfnamefont {K.~H.}\ \bibnamefont
  {Nam}},\ }\href@noop {} {\bibfield  {journal} {\bibinfo  {journal} {J. Math.
  Anal. App.}\ }\textbf {\bibinfo {volume} {406}},\ \bibinfo {pages} {101}
  (\bibinfo {year} {2013})}\BibitemShut {NoStop}%
\bibitem [{\citenamefont {Spruck}\ and\ \citenamefont
  {Yang}(1992{\natexlab{a}})}]{Spruck1}%
  \BibitemOpen
  \bibfield  {author} {\bibinfo {author} {\bibfnamefont {J.}~\bibnamefont
  {Spruck}}\ and\ \bibinfo {author} {\bibfnamefont {Y.}~\bibnamefont {Yang}},\
  }\href@noop {} {\bibfield  {journal} {\bibinfo  {journal} {Commun. Math.
  Phys.}\ }\textbf {\bibinfo {volume} {144}},\ \bibinfo {pages} {1} (\bibinfo
  {year} {1992}{\natexlab{a}})}\BibitemShut {NoStop}%
\bibitem [{\citenamefont {Spruck}\ and\ \citenamefont
  {Yang}(1992{\natexlab{b}})}]{Spruck2}%
  \BibitemOpen
  \bibfield  {author} {\bibinfo {author} {\bibfnamefont {J.}~\bibnamefont
  {Spruck}}\ and\ \bibinfo {author} {\bibfnamefont {Y.}~\bibnamefont {Yang}},\
  }\href@noop {} {\bibfield  {journal} {\bibinfo  {journal} {Commun. Math.
  Phys.}\ }\textbf {\bibinfo {volume} {144}},\ \bibinfo {pages} {215} (\bibinfo
  {year} {1992}{\natexlab{b}})}\BibitemShut {NoStop}%
\bibitem [{\citenamefont {Spruck}\ and\ \citenamefont {Yang}(1995)}]{Spruck3}%
  \BibitemOpen
  \bibfield  {author} {\bibinfo {author} {\bibfnamefont {J.}~\bibnamefont
  {Spruck}}\ and\ \bibinfo {author} {\bibfnamefont {Y.}~\bibnamefont {Yang}},\
  }\href@noop {} {\bibfield  {journal} {\bibinfo  {journal} {H.
  Poincar\'{e}-Anal. non lin\'{e}aire}\ }\textbf {\bibinfo {volume} {12}},\
  \bibinfo {pages} {75} (\bibinfo {year} {1995})}\BibitemShut {NoStop}%
\bibitem [{\citenamefont {Spruck}\ and\ \citenamefont
  {Yang}(1992{\natexlab{c}})}]{Spruck4}%
  \BibitemOpen
  \bibfield  {author} {\bibinfo {author} {\bibfnamefont {J.}~\bibnamefont
  {Spruck}}\ and\ \bibinfo {author} {\bibfnamefont {Y.}~\bibnamefont {Yang}},\
  }\href@noop {} {\bibfield  {journal} {\bibinfo  {journal} {Commun. Math.
  Phys.}\ }\textbf {\bibinfo {volume} {149}},\ \bibinfo {pages} {361} (\bibinfo
  {year} {1992}{\natexlab{c}})}\BibitemShut {NoStop}%
\bibitem [{\citenamefont {Bartolucci}\ and\ \citenamefont
  {Tarantello}(2002)}]{Tarantello1}%
  \BibitemOpen
  \bibfield  {author} {\bibinfo {author} {\bibfnamefont {D.}~\bibnamefont
  {Bartolucci}}\ and\ \bibinfo {author} {\bibfnamefont {G.}~\bibnamefont
  {Tarantello}},\ }\href@noop {} {\bibfield  {journal} {\bibinfo  {journal}
  {Comm. Math. Phys.}\ }\textbf {\bibinfo {volume} {229}},\ \bibinfo {pages}
  {3} (\bibinfo {year} {2002})}\BibitemShut {NoStop}%
\bibitem [{\citenamefont {Tarantello}(2010)}]{Tarantello2}%
  \BibitemOpen
  \bibfield  {author} {\bibinfo {author} {\bibfnamefont {G.}~\bibnamefont
  {Tarantello}},\ }\href@noop {} {\bibfield  {journal} {\bibinfo  {journal}
  {Discrete Contin. Dyn. Syst.}\ }\textbf {\bibinfo {volume} {28}},\ \bibinfo
  {pages} {931} (\bibinfo {year} {2010})}\BibitemShut {NoStop}%
\bibitem [{\citenamefont {Tarantello}(2008)}]{Tarantello3}%
  \BibitemOpen
  \bibfield  {author} {\bibinfo {author} {\bibfnamefont {G.}~\bibnamefont
  {Tarantello}},\ }\href@noop {} {\emph {\bibinfo {title} {Self-Dual Gauge
  Field Vortices. Progress in Nonlinear Differential Equations and Their
  Applications}}}\ (\bibinfo  {publisher} {Birkauser},\ \bibinfo {year}
  {2008})\BibitemShut {NoStop}%
\bibitem [{\citenamefont {Wang}\ and\ \citenamefont {Yang}(1992)}]{Wang}%
  \BibitemOpen
  \bibfield  {author} {\bibinfo {author} {\bibfnamefont {S.}~\bibnamefont
  {Wang}}\ and\ \bibinfo {author} {\bibfnamefont {Y.}~\bibnamefont {Yang}},\
  }\href@noop {} {\bibfield  {journal} {\bibinfo  {journal} {SIAM J. Math.
  Anal.}\ }\textbf {\bibinfo {volume} {23}},\ \bibinfo {pages} {1125} (\bibinfo
  {year} {1992})}\BibitemShut {NoStop}%
\bibitem [{\citenamefont {Yang}(1997)}]{Yang1}%
  \BibitemOpen
  \bibfield  {author} {\bibinfo {author} {\bibfnamefont {Y.}~\bibnamefont
  {Yang}},\ }\href@noop {} {\bibfield  {journal} {\bibinfo  {journal} {Physica
  D}\ }\textbf {\bibinfo {volume} {101}},\ \bibinfo {pages} {55} (\bibinfo
  {year} {1997})}\BibitemShut {NoStop}%
\bibitem [{\citenamefont {Yang}(2000{\natexlab{a}})}]{Yang2}%
  \BibitemOpen
  \bibfield  {author} {\bibinfo {author} {\bibfnamefont {Y.}~\bibnamefont
  {Yang}},\ }\href@noop {} {\bibfield  {journal} {\bibinfo  {journal} {Proc.
  Roy. Soc. A}\ }\textbf {\bibinfo {volume} {456}},\ \bibinfo {pages} {615}
  (\bibinfo {year} {2000}{\natexlab{a}})}\BibitemShut {NoStop}%
\bibitem [{\citenamefont {Yang}(2000{\natexlab{b}})}]{Yang3}%
  \BibitemOpen
  \bibfield  {author} {\bibinfo {author} {\bibfnamefont {Y.}~\bibnamefont
  {Yang}},\ }\href@noop {} {\bibfield  {journal} {\bibinfo  {journal} {J.
  Funct. Anal.}\ }\textbf {\bibinfo {volume} {170}},\ \bibinfo {pages} {1}
  (\bibinfo {year} {2000}{\natexlab{b}})}\BibitemShut {NoStop}%
\bibitem [{\citenamefont {Yang}(2001)}]{Yang4}%
  \BibitemOpen
  \bibfield  {author} {\bibinfo {author} {\bibfnamefont {Y.}~\bibnamefont
  {Yang}},\ }\href@noop {} {\emph {\bibinfo {title} {Solitons in Field Theory
  and Nonlinear Analysis}}}\ (\bibinfo  {publisher} {Springer Monographs in
  Mathematics},\ \bibinfo {year} {2001})\BibitemShut {NoStop}%
\bibitem [{\citenamefont {Chakraborty}\ and\ \citenamefont
  {Pietilainen}(1995)}]{Cha}%
  \BibitemOpen
  \bibfield  {author} {\bibinfo {author} {\bibfnamefont {T.}~\bibnamefont
  {Chakraborty}}\ and\ \bibinfo {author} {\bibfnamefont {P.}~\bibnamefont
  {Pietilainen}},\ }\href@noop {} {\emph {\bibinfo {title} {The Quantum Hall
  Effects}}}\ (\bibinfo  {publisher} {Springer},\ \bibinfo {year}
  {1995})\BibitemShut {NoStop}%
\bibitem [{\citenamefont {Frolich}(1995)}]{Frolich1}%
  \BibitemOpen
  \bibfield  {author} {\bibinfo {author} {\bibfnamefont {J.}~\bibnamefont
  {Frolich}},\ }\href@noop {} {\bibfield  {journal} {\bibinfo  {journal} {Proc.
  Internat. Congr. Math., Birkhauser}\ ,\ \bibinfo {pages} {75}} (\bibinfo
  {year} {1995})}\BibitemShut {NoStop}%
\bibitem [{\citenamefont {Frolich}\ and\ \citenamefont
  {Marchetti}(1988)}]{Frolich2}%
  \BibitemOpen
  \bibfield  {author} {\bibinfo {author} {\bibfnamefont {J.}~\bibnamefont
  {Frolich}}\ and\ \bibinfo {author} {\bibfnamefont {P.}~\bibnamefont
  {Marchetti}},\ }\href@noop {} {\bibfield  {journal} {\bibinfo  {journal}
  {Lett. Math. Phys.}\ }\textbf {\bibinfo {volume} {16}},\ \bibinfo {pages}
  {347} (\bibinfo {year} {1988})}\BibitemShut {NoStop}%
\bibitem [{\citenamefont {Frolich}\ and\ \citenamefont
  {Marchetti}(1989)}]{Frolich3}%
  \BibitemOpen
  \bibfield  {author} {\bibinfo {author} {\bibfnamefont {J.}~\bibnamefont
  {Frolich}}\ and\ \bibinfo {author} {\bibfnamefont {P.}~\bibnamefont
  {Marchetti}},\ }\href@noop {} {\bibfield  {journal} {\bibinfo  {journal}
  {Commun. Math. Phys.}\ }\textbf {\bibinfo {volume} {121}},\ \bibinfo {pages}
  {177} (\bibinfo {year} {1989})}\BibitemShut {NoStop}%
\bibitem [{\citenamefont {Girvin}(2000)}]{Girvin1}%
  \BibitemOpen
  \bibfield  {author} {\bibinfo {author} {\bibfnamefont {S.~M.}\ \bibnamefont
  {Girvin}},\ }\href@noop {} {\bibfield  {journal} {\bibinfo  {journal} {Les
  Houches lectures}\ }\textbf {\bibinfo {volume} {29}},\ \bibinfo {pages} {53}
  (\bibinfo {year} {2000})}\BibitemShut {NoStop}%
\bibitem [{\citenamefont {Girvin}\ and\ \citenamefont
  {Prange}(1990)}]{Girvin2}%
  \BibitemOpen
  \bibfield  {author} {\bibinfo {author} {\bibfnamefont {S.~M.}\ \bibnamefont
  {Girvin}}\ and\ \bibinfo {author} {\bibfnamefont {R.~E.}\ \bibnamefont
  {Prange}},\ }\href@noop {} {\emph {\bibinfo {title} {The Quantum Hall
  Effect}}},\ \bibinfo {edition} {2nd}\ ed.\ (\bibinfo  {publisher}
  {Springer},\ \bibinfo {year} {1990})\BibitemShut {NoStop}%
\bibitem [{\citenamefont {Ichinose}\ and\ \citenamefont
  {Sekiguchi}(1997{\natexlab{a}})}]{Ichinose1}%
  \BibitemOpen
  \bibfield  {author} {\bibinfo {author} {\bibfnamefont {I.}~\bibnamefont
  {Ichinose}}\ and\ \bibinfo {author} {\bibfnamefont {A.}~\bibnamefont
  {Sekiguchi}},\ }\href@noop {} {\bibfield  {journal} {\bibinfo  {journal}
  {Mod. Phys. Lett. A}\ }\textbf {\bibinfo {volume} {12}},\ \bibinfo {pages}
  {2243} (\bibinfo {year} {1997}{\natexlab{a}})}\BibitemShut {NoStop}%
\bibitem [{\citenamefont {Ichinose}\ and\ \citenamefont
  {Sekiguchi}(1997{\natexlab{b}})}]{Ichinose2}%
  \BibitemOpen
  \bibfield  {author} {\bibinfo {author} {\bibfnamefont {I.}~\bibnamefont
  {Ichinose}}\ and\ \bibinfo {author} {\bibfnamefont {A.}~\bibnamefont
  {Sekiguchi}},\ }\href@noop {} {\bibfield  {journal} {\bibinfo  {journal}
  {Nuclear Physics B}\ }\textbf {\bibinfo {volume} {492}},\ \bibinfo {pages}
  {683} (\bibinfo {year} {1997}{\natexlab{b}})}\BibitemShut {NoStop}%
\bibitem [{\citenamefont {Jackiw}\ and\ \citenamefont {Pi}(1990)}]{Jackiw}%
  \BibitemOpen
  \bibfield  {author} {\bibinfo {author} {\bibfnamefont {R.}~\bibnamefont
  {Jackiw}}\ and\ \bibinfo {author} {\bibfnamefont {S.-Y.}\ \bibnamefont
  {Pi}},\ }\href@noop {} {\bibfield  {journal} {\bibinfo  {journal} {Phys. Rev.
  Lett.}\ }\textbf {\bibinfo {volume} {64}},\ \bibinfo {pages} {29} (\bibinfo
  {year} {1990})}\BibitemShut {NoStop}%
\bibitem [{\citenamefont {Jain}(1989)}]{Jain}%
  \BibitemOpen
  \bibfield  {author} {\bibinfo {author} {\bibfnamefont {J.~K.}\ \bibnamefont
  {Jain}},\ }\href@noop {} {\bibfield  {journal} {\bibinfo  {journal} {Phys.
  Rev. Lett.}\ }\textbf {\bibinfo {volume} {63}},\ \bibinfo {pages} {199}
  (\bibinfo {year} {1989})}\BibitemShut {NoStop}%
\bibitem [{\citenamefont {v.~Klitzing}(1984)}]{Klitzing}%
  \BibitemOpen
  \bibfield  {author} {\bibinfo {author} {\bibfnamefont {K.}~\bibnamefont
  {v.~Klitzing}},\ }\href@noop {} {\bibfield  {journal} {\bibinfo  {journal}
  {Physica B+C}\ }\textbf {\bibinfo {volume} {126}},\ \bibinfo {pages} {242}
  (\bibinfo {year} {1984})}\BibitemShut {NoStop}%
\bibitem [{\citenamefont {Kohmoto}(1985)}]{Kohmoto}%
  \BibitemOpen
  \bibfield  {author} {\bibinfo {author} {\bibfnamefont {M.}~\bibnamefont
  {Kohmoto}},\ }\href@noop {} {\bibfield  {journal} {\bibinfo  {journal}
  {Annals of Physics}\ }\textbf {\bibinfo {volume} {160}},\ \bibinfo {pages}
  {343} (\bibinfo {year} {1985})}\BibitemShut {NoStop}%
\bibitem [{\citenamefont {McDonald}(1990)}]{McDonald}%
  \BibitemOpen
  \bibfield  {author} {\bibinfo {author} {\bibfnamefont {A.~H.}\ \bibnamefont
  {McDonald}},\ }\href@noop {} {\emph {\bibinfo {title} {Quantum Hall Effect: A
  perspective}}}\ (\bibinfo  {publisher} {Kluwer Academic Publishing},\
  \bibinfo {year} {1990})\BibitemShut {NoStop}%
\bibitem [{\citenamefont {Stone}(1992)}]{Stone}%
  \BibitemOpen
  \bibfield  {author} {\bibinfo {author} {\bibfnamefont {M.}~\bibnamefont
  {Stone}},\ }\href@noop {} {\emph {\bibinfo {title} {Quantum Hall Effect}}}\
  (\bibinfo  {publisher} {World Scientific},\ \bibinfo {year}
  {1992})\BibitemShut {NoStop}%
\bibitem [{\citenamefont {Thoules}(1984{\natexlab{a}})}]{Thoules1}%
  \BibitemOpen
  \bibfield  {author} {\bibinfo {author} {\bibfnamefont {D.}~\bibnamefont
  {Thoules}},\ }\href@noop {} {\bibfield  {journal} {\bibinfo  {journal}
  {Surface Science}\ }\textbf {\bibinfo {volume} {142}},\ \bibinfo {pages}
  {147} (\bibinfo {year} {1984}{\natexlab{a}})}\BibitemShut {NoStop}%
\bibitem [{\citenamefont {Thoules}(1984{\natexlab{b}})}]{Thoules2}%
  \BibitemOpen
  \bibfield  {author} {\bibinfo {author} {\bibfnamefont {D.}~\bibnamefont
  {Thoules}},\ }\href@noop {} {\bibfield  {journal} {\bibinfo  {journal}
  {Physical Reports}\ }\textbf {\bibinfo {volume} {110}},\ \bibinfo {pages}
  {279} (\bibinfo {year} {1984}{\natexlab{b}})}\BibitemShut {NoStop}%
\bibitem [{\citenamefont {Bogomol'nyi}(1976)}]{Bogo}%
  \BibitemOpen
  \bibfield  {author} {\bibinfo {author} {\bibfnamefont {E.~B.}\ \bibnamefont
  {Bogomol'nyi}},\ }\href@noop {} {\bibfield  {journal} {\bibinfo  {journal}
  {Sov. J. Nucl. Phys.}\ }\textbf {\bibinfo {volume} {24}},\ \bibinfo {pages}
  {449} (\bibinfo {year} {1976})}\BibitemShut {NoStop}%
\bibitem [{\citenamefont {Prasad}\ and\ \citenamefont
  {Sommerfield}(1975)}]{Prasad}%
  \BibitemOpen
  \bibfield  {author} {\bibinfo {author} {\bibfnamefont {M.~K.}\ \bibnamefont
  {Prasad}}\ and\ \bibinfo {author} {\bibfnamefont {C.}~\bibnamefont
  {Sommerfield}},\ }\href@noop {} {\bibfield  {journal} {\bibinfo  {journal}
  {Phys. Rev. Lett.}\ }\textbf {\bibinfo {volume} {35}},\ \bibinfo {pages}
  {760} (\bibinfo {year} {1975})}\BibitemShut {NoStop}%
\bibitem [{\citenamefont {'t~Hooft}(1979)}]{Hooft}%
  \BibitemOpen
  \bibfield  {author} {\bibinfo {author} {\bibfnamefont {G.}~\bibnamefont
  {'t~Hooft}},\ }\href@noop {} {\bibfield  {journal} {\bibinfo  {journal}
  {Nucl. Phys. B}\ }\textbf {\bibinfo {volume} {153}},\ \bibinfo {pages} {141}
  (\bibinfo {year} {1979})}\BibitemShut {NoStop}%
\bibitem [{\citenamefont {Aubin}(1982)}]{Aubin}%
  \BibitemOpen
  \bibfield  {author} {\bibinfo {author} {\bibfnamefont {T.}~\bibnamefont
  {Aubin}},\ }\href@noop {} {\emph {\bibinfo {title} {Nonlinear Analysis on
  Manifolds in Monge-Ampere Equations}}}\ (\bibinfo  {publisher} {Springer},\
  \bibinfo {year} {1982})\BibitemShut {NoStop}%
\bibitem [{\citenamefont {Jaffe}\ and\ \citenamefont {Taubes}(1980)}]{Taubes}%
  \BibitemOpen
  \bibfield  {author} {\bibinfo {author} {\bibfnamefont {A.}~\bibnamefont
  {Jaffe}}\ and\ \bibinfo {author} {\bibfnamefont {C.~H.}\ \bibnamefont
  {Taubes}},\ }\href@noop {} {\emph {\bibinfo {title} {Vortices and
  Monopoles}}}\ (\bibinfo  {publisher} {Birkhauser},\ \bibinfo {year}
  {1980})\BibitemShut {NoStop}%
\end{thebibliography}%
% ------------------------------------------------------------------------
\end{document}